\documentclass[iop]{emulateapj}

\usepackage{natbib}

\shorttitle{Identifying high-redshift GRBs with RATIR}
\shortauthors{Littlejohns et al.}

\begin{document}

\title{Identifying high-redshift GRBs with RATIR}

\author{O.~M.~Littlejohns\altaffilmark{1}, N.~R.~Butler\altaffilmark{1}, A.~Cucchiara\altaffilmark{2}, A.~M.~Watson\altaffilmark{3}, A.~S.~Kutyrev\altaffilmark{4}, W.~H.~Lee\altaffilmark{3}, M.~G.~Richer\altaffilmark{3}, C.~R.~Klein\altaffilmark{5}, O.~D.~Fox\altaffilmark{5}, J.~X.~Prochaska\altaffilmark{6}, J.~S.~Bloom\altaffilmark{5}, E.~Troja\altaffilmark{4}, E.~Ramirez-Ruiz\altaffilmark{6}, J.~A.~de~Diego\altaffilmark{3}, L.~Georgiev\altaffilmark{3}, J.~Gonz\'{a}lez\altaffilmark{3}, C.~G.~Rom\'{a}n-Z\'{u}\~{n}iga\altaffilmark{3}, N.~Gehrels\altaffilmark{4} and H.~Moseley\altaffilmark{4}}
\altaffiltext{1}{School of Earth \& Space Exploration, Arizona State University,
    AZ 85287, USA}
\altaffiltext{2}{NASA Postdoctoral Program Fellow, Goddard Space Flight Center, Greenbelt, MD 20771, USA}
\altaffiltext{3}{Instituto de Astronom\'{i}a, Universidad Nacional Aut\'{o}noma de M\'{e}xico, Apartado Postal 70-264, 04510 M\'{e}xico, D. F., M\'{e}xico}
\altaffiltext{4}{NASA, Goddard Space Flight Center, Greenbelt, MD 20771, USA}
\altaffiltext{5}{Astronomy Department, University of California, Berkeley,
    CA 94720-7450, USA}
\altaffiltext{6}{Department of Astronomy and Astrophysics, UCO/Lick Observatory, University of California, 1156 High Street, Santa Cruz, CA 95064, USA}

\begin{abstract}
We present  a template  fitting algorithm for  determining photometric
redshifts, $z_{\rm phot}$, of candidate high-redshift gamma-ray bursts
(GRBs). Using  afterglow photometry, obtained by  the Reionization And
Transients InfraRed  (RATIR) camera,  this algorithm accounts  for the
intrinsic GRB afterglow spectral  energy distribution (SED), host dust
extinction and the effect of neutral hydrogen (local and cosmological)
along  the line of  sight.  We  present the  results obtained  by this
algorithm and RATIR photometry of GRB~130606A, finding a range of best
fit solutions  $5.6 < z_{\rm phot}  < 6.0$ for models  of several host
dust  extinction  laws  (none,  MW,  LMC  and  SMC),  consistent  with
spectroscopic  measurements  of  the  redshift  of  this  GRB.   Using
simulated  RATIR photometry,  we find  our algorithm  provides precise
measures of $z_{\rm phot}$ in the ranges $4 < z_{\rm phot} \lesssim 8$
and $9  < z_{\rm phot} <  10$ and can robustly  determine when $z_{\rm
  phot}>4$.  Further testing highlights  the required caution in cases
of highly  dust extincted host  galaxies.  These tests also  show that
our  algorithm does  not  erroneously  find $z_{\rm  phot}  < 4$  when
$z_{\rm sim}>4$, thereby minimizing false negatives and allowing us to
rapidly identify all potential high-redshift events.
\end{abstract}


\section{Introduction}

Gamma-ray                         bursts                        (GRBs)
\citep{2009ARA&A..47..567G,2006RPPh...69.2259M}   are  bright  objects
that emit across the entire electromagnetic spectrum and therefore can
be  seen to  high-redshift  \citep{2000ApJ...536....1L}.  This  allows
observers  to identify  the  locations of  high-redshift, faint,  host
galaxy population otherwise missed by the current and future magnitude
limited  surveys \citep{2012ApJ...754...46T}.   Such  discoveries will
allow  us to  unveil the  properties of  early star  formation  (up to
$z\sim10$) and the epoch of reionization.\par

To date a  number of high-redshift GRBs (all  in the redshift interval
$6<   z   <   10$)   have   been   identified   including   GRB~050904
\citep{2006PASJ...58..485T},  GRB~080913  \citep{2009ApJ...693.1610G},
GRB~090423       \citep{2009Natur.461.1254T}      and      GRB~090429B
\citep{2011ApJ...736....7C}.   However, it  is only  in a  more recent
example,  GRB~130606A at  $z=5.913$  \citep{2013ApJ...774...26C}, that
high  signal  to   noise  ratio  spectrum  has  been   obtained  of  a
high-redshift GRB afterglow.  Using  a method similar to that employed
for   quasars   \citep{2006AJ....132..117F,2001AJ....122.2850B},   the
detection of  Ly$\beta$ and  Ly$\gamma$ transmission in  this spectrum
allowed \citet{2013ApJ...774...26C} to identify that the intergalactic
medium (IGM) is mostly ionized at this redshift, and that the epoch of
reionization  must  have  occurred  earlier  in  the  history  of  the
Universe.\par

\citet{2013arXiv1312.5631C}  also  present  broadband photometric  and
spectroscopic observations  of GRB~130606A, finding a  lower HI column
density     of     $N_{\rm    HI}     =     7.1    \times     10^{19}$
cm$^{-2}$. \citet{2013arXiv1312.3934T} consider a possible intervening
damped  Lyman-$\alpha$  (DLA) system  at  $z_{\rm  DLA}  =5.806$ as  a
contributor  to  the absorption  via  HI  gas.  Considerations of  the
required   silicon    abundance   in   such   a    DLA   system   lead
\citet{2013arXiv1312.3934T}  to  instead  attribute the  residuals  in
their host-only model to diffuse IGM  along the line of sight at $5.83
< z_{\rm IGM}< 5.91$.\par

To   obtain   high   signal-to-noise   ratio  spectra   of   potential
high-redshift GRB,  afterglows must first be  rapidly identified.  The
primary  method  used to  do  so  is  the measurement  of  photometric
redshifts from  optical ``dropout'' candidates.   As radiation travels
through  the  Universe,  clouds  of  neutral  hydrogen  attenuate  the
transmitted radiation \citep{1995ApJ...441...18M}.  This effect occurs
along  the line  of sight  between source  and observer,  and  so this
attenuation encroaches into increasingly  redder parts of the observed
optical and  near infrared (NIR)  spectrum for sources  (and therefore
atomic  hydrogen  clouds)  at  higher  redshifts.\par

In practice,  few facilities can  provide the necessary  photometry to
identify high-redshift candidate as most automatic, robotic telescopes
use optical  filters and  as such are  not sensitive  to high-redshift
GRBs.  In some instances, where large aperture ground-based facilities
suffer  observational  constraints, spectral  information  may not  be
obtained  prior  to  the   GRB  afterglow  fading  below  instrumental
sensitivity       limits,      such      as       for      GRB~090429B
\citep{2011ApJ...736....7C},  providing further  motivation  to obtain
photometric  redshift,  $z_{\rm  phot}$, estimates  for  high-redshift
dropout candidates.   Whilst $z_{\rm phot}=9.38_{-0.32}^{+0.14}$ (90\%
likelihood range)  was later obtained,  additional rapid-response from
GRB-dedicated NIR  multiple filter instruments would  have assisted in
providing  a measure  of $z_{\rm  phot}$ and  allowing  large aperture
telescopes to begin observations  at the earliest possible epoch after
the initial trigger.\par

The variable $z_{\rm  phot}$ is estimated by fitting  templates to the
optical and  NIR spectral energy distributions (SEDs;  see for example
\citealt{2011A&A...526A.153K}    and    \citealt{2008A&A...490.1047C})
observed  for  the  afterglow  of  the GRB,  which  has  an  intrinsic
synchrotron                                                    spectrum
\citep{2002ApJ...568..820G,1998ApJ...497L..17S}. Through this template
fitting method the redshift of  a GRB, the host galaxy dust extinction
and  the  optical/NIR spectral  index  of  the  GRB afterglow  can  be
estimated.   Fitting a  host  extinction  law allows  a  study of  the
prevalence  of   dust  in  early   galaxies  along  GRB   sight  lines
\citep{2011ApJ...735....2Z},  whilst the local  spectral index  can be
compared  to that  obtained in  other energy  ranges to  determine key
parameters     of      the     underlying     synchrotron     spectrum
\citep{2013arXiv1307.4401P}.\par

To   determine   $z_{\rm   phot}$,  \citet{2011A&A...526A.153K}   make
extensive use  of data from the  Gamma-Ray Burst Optical/Near-Infrared
Detector (GROND; \citealt{2008PASP..120..405G}).  GROND simultaneously
images  in  seven   filters:  \textit{g'},  \textit{r'},  \textit{i'},
\textit{z'},   \textit{J},   \textit{H}   and  \textit{K}$_{\rm   S}$,
providing  a broad  optical and  NIR  SED.  Such  an SED  was used  to
determine    the    $z_{\rm    phot}=4.35\pm0.15$   for    GRB~080916C
\citep{2009A&A...498...89G} and another indicated helped identify that
GRB~080913 was a high redshift event ($z=6.695\pm0.025$) and warranted
further     observations     from     large    aperture     facilities
\citep{2009ApJ...693.1610G}.\par

Another instrument  capable of providing  the photometric observations
needed to determine $z_{\rm  phot}$ is the Reionization And Transients
Infra-Red  (RATIR;  \citealt{2012SPIE.8446E..10B})  camera,  which  is
mounted  on  the  1.5-meter   Harold  L.   Johnson  telescope  of  the
Observatorio Astron\'{o}mico  Nacional on Sierra  San Pedro M\'{a}rtir
in  Baja  California,  Mexico.   This  facility,  which  became  fully
operational  in   December  2012,  conducts   autonomous  observations
\citep{2012SPIE.8444E..5LW,2012SPIE.8453E..2SK}  of GRB  triggers from
the \textit{Swift} satellite \citep{2004ApJ...611.1005G}.  Since RATIR
obtains simultaneous photometry in \textit{r}, \textit{i}, \textit{Z},
\textit{Y}, \textit{J}  and \textit{H}, it is  an excellent instrument
for  estimating $z_{\rm  phot}$  for high-redshift  GRBs  and as  such
allows   us  to  optimize   spectroscopic  observations   with  highly
oversubscribed large telescopes.\par

RATIR and GROND  operate in similar manners, both  with best responses
of order a  few minutes after the initial  \textit{Swift} trigger time
\citep{2013GCN..14943...1B,2010GCN..10874...1U}.  The filters employed
by  both instruments  are similar,  but  not identical.   GROND has  a
broader  spectral  coverage,  extending   to  both  higher  and  lower
wavelengths. The lower wavelength  coverage aids in the identification
of $2.3 <  z_{\rm phot} < 4$, although  RATIR is specifically designed
to target objects in the high-redshift Universe.  While GROND also has
a  \textit{K}$_{\rm  S}$-band  filter,  extending coverage  to  higher
redshifts, RATIR  contains a \textit{Y}-band filter  that GROND lacks.
With such functionality, RATIR is  better able to provide more precise
estimates of $z_{\rm  phot}$ when a GRB  occurs $7 < z <  8$. RATIR is
also  100\%  time  dedicated  to  GRB  follow-up.   Perhaps  the  most
important  distinction  between  the   two  instruments  is  in  their
locations.     RATIR    is    situated    at    latitude    $\phi    =
+31^{\circ}02^{\prime}43^{\prime\prime}$,  while GROND  is  mounted on
the 2.2-meter MPI/ESO telescope at La Silla Observatory, Chile, with a
latitude $\phi = -29^{\circ}15^{\prime}15^{\prime\prime}$, meaning the
both  form a  complimentary  pair of  instruments routinely  accessing
different parts of the sky.\par

This work  describes the development of our  template fitting routine.
In \S \ref{sec:method} we describe  the models employed to produce the
templates  fitted  to   the  RATIR  optical  and  NIR   data.   In  \S
\ref{sec:obs} we describe the \textit{Swift} and RATIR observations of
GRB~130606A,  before in  \S \ref{sec:130606A}  presenting  the results
from analysis of the early epoch RATIR data.  We then extensively test
the  capabilities of  our algorithm  with simulated  RATIR data  in \S
\ref{sec:tests}.    Finally,  we   present  our   conclusions   in  \S
\ref{sec:conc}\par

\section{Method}
\label{sec:method}

To  measure $z_{\rm  phot}$ we  adopt a  template  fitting methodology
similar to that of  \citet{2011A&A...526A.153K}, who use combined data
sets  from GROND  and  both the  Ultraviolet/Optical Telescope  (UVOT;
\citealt{2005SSRv..120...95R})     and    X-ray     Telescope    (XRT;
\citealt{2005SSRv..120..165B}) on  board the \textit{Swift} satellite.
Another  similar methodology  is that  of \citet{2008A&A...490.1047C},
although   the   algorithm   we   implement  is   significantly   less
computationally  expensive.   To  this   end  we  use  models  of  the
underlying physical spectrum of the source and the mechanisms by which
this spectrum is altered before  reaching an observer at Earth.  These
processes are  dust absorption from the host  galaxy, attenuation from
the Intergalactic Medium  (IGM) and the response of  the RATIR filters
to incoming photons.\par

Unlike  \citet{2011A&A...526A.153K} our  algorithm  does not  consider
UVOT  photometry, as  only approximately  26\% of  \textit{Swift} GRBs
having UVOT detections, compared  to the $\sim$60\% of bursts detected
by  ground-based  facilities  \citep{2009ApJ...690..163R}.   Following
\citet{2012A&A...537A..15S},  we  also favor  the  extinction laws  of
\citet{1992ApJ...395..130P},  instead of  the  more general  ``Drude''
model           \citep{2009ApJ...690L..56L}           used          by
\citet{2011A&A...526A.153K}. This was done  in an attempt to limit the
number  of parameters  being fitted  as the  ``Drude'' model  has four
parameters  compared  to  the  one  ($A_{V}$)  used  in  the  standard
\citet{1992ApJ...395..130P} extinction laws.\par

\subsection{Intrinsic GRB spectrum}
\label{sec:intspec}

The   emission  of   GRBs  is   attributed  to   synchrotron  emission
\citep{1998ApJ...497L..17S}, which produces a spectrum consisting of a
series of broken power-laws.   The location of the synchrotron cooling
break, usually  found between the  optical and X-ray  regimes, evolves
with  time in a  manner determined  by the  nature of  the circumburst
medium  \citep{2002ApJ...568..820G}.  The passage  of  such a  cooling
break through the optical and  NIR bands would manifest itself through
an achromatic temporal  break in the GRB light curve,  which we do not
observe for GRB~130606A.   Due to this, and to  avoid the complication
of additional fitting  parameters, we assume that the  regime in which
RATIR observes was entirely within only one of the power-law segments,
thus:

\begin{equation}
  F \left( \lambda \right) = F_{0} \left( \frac{\lambda}{\lambda_{0}} 
  \right)^{\beta_{{\rm opt}}},
  \label{eq:intspec}
\end{equation}

where $F\left( \lambda  \right)$ is flux as a  function of wavelength,
$\lambda$, $F_{0}$ is the normalization corresponding to the flux at a
designated wavelength,  $\lambda_{0}$, and $\beta_{{\rm  opt}}$ is the
local   spectral  index   of  the   power-law.   Both   $\lambda$  and
$\lambda_{0}$  are  in  the  rest  frame  of  the  radiating  outflow.
Traditionally, the spectrum  of a GRB is represented  as a function of
frequency rather  that wavelength, however the  expression in Equation
\ref{eq:intspec} is entirely equivalent and easier to combine with the
dust  extinction  and  hydrogen   absorption  models  detailed  in  \S
\ref{sec:dustabs} and \S \ref{sec:IGMatten}.\par

\subsection{Host dust absorption}
\label{sec:dustabs}

Once emitted, the first external  effect to act upon the GRB radiation
is absorption by  dust in the medium of the host  galaxy along the GRB
sight line.   Most GRB host galaxies  detected in the  optical and NIR
regimes exhibit low  amounts of dust extinction along  the GRB line of
sight    \citep{2007ApJ...661..787S,2006ApJ...641..993K}    with   the
majority of SED models for a  sample covering $0.1 < z < 4.5$ favoring
$A_{V}\leqslant 0.2$.  NIR data have revealed a subset ($\sim10$\%) of
GRBs to be highly extinguished ($A_{V}\gtrsim 1$) by the dust in their
host galaxy  from a  sample spanning $0.5  \leqslant z  \leqslant 9.4$
\citep{2011A&A...526A..30G}.   \citet{2012A&A...537A..15S}  model  the
extinction  laws  for  a  sample  of  GRB  host  galaxies  showing  an
ultraviolet  (UV) slope  comparable to  that of  the  Large Magellanic
Clouds (LMC)  but an absence of  the 2175~\AA\ ``bump''  in hosts with
low   dust    extinction.    Conversely,   \citet{2012A&A...537A..15S}
demonstrate  that  hosts  of   highly  extinguished  GRBs  show  clear
detections  of  the  2175~\AA\  bump and  flatter  extinction  curves,
thereby are more consistent with that of the MW.\par

To   account  for   host   galaxy  dust   extinction,   we  used   the
graphite-silicate      dust     grain     models      presented     in
\citet{1992ApJ...395..130P}.   As the  dust extinction  of a  GRB host
galaxy along  the line  of sight is  not known \textit{a  priori}, the
template  algorithm  implements  models  for  the MW,  LMC  and  Small
Magellanic Clouds  (SMC).  We also consider  a case with  no host dust
absorption.\par

To calculate dust  absorption as a function of  wavelength, we use the
empirical    extinction    curves   presented    in    Table   1    of
\citet{1992ApJ...395..130P} and  the corresponding values  of $R_{V}$,
where  $R_{V}\equiv A_{V}/E_{B-V}$  and $A_{V}$  is the  extinction in
magnitudes   of  the   $V$-band.    $E_{B-V}=A_{B}-A_{V}$  gives   the
difference in extinction between  the $B$- and $V$-bands. Both $A_{V}$
and  $E_{B-V}$  are defined  in  terms  of  rest-frame \textit{B}  and
\textit{V}.\par

We  dust   extinction  at  each   wavelength  given  in  Table   1  of
\citet{1992ApJ...395..130P}, using Equation \ref{eq:peiext}, below. We
then re-sample the extinction law on to a finer grid.\par

\begin{equation}
  A_{\lambda} = A_{V} \left( \frac{1}{R_{V}}\frac{E_{\lambda-V}}{E_{B-V}} + 1 \right).
  \label{eq:peiext}
\end{equation}

\subsection{IGM attenuation}
\label{sec:IGMatten}

The algorithm then calculates attenuation from the intervening IGM due
to absorption from clouds of  neutral hydrogen along the line of sight
\citep{1965ApJ...142.1633G}.  We  adopt a similar  methodology to that
used                   in                   the                   {\sc
  hyperz}\footnote{http://webast.ast.obs-mip.fr/hyperz/} software (see
also  \citealt{2000A&A...363..476B}) by  estimating  the reduction  of
flux   from  Ly$\alpha$  and   Ly$\beta$  absorption   using  Equation
\ref{eq:dfac},   below,   which   is   based   on   Equation   17   of
\citet{1995ApJ...441...18M}:\par

\tiny
\begin{equation}
  D \left( \lambda \right) = \left\{
  \begin{array}{ll}
    1, & \mbox{$\lambda_{\rm rest} > 1216$ \AA,} \\
    \frac{1}{\Delta \lambda_{A}} \times & \\[4ex]
    \int\limits_{1050\left(1+z\right)}^{1170\left(1+z\right)}
    \exp \left[ - A_{2} \left( \frac{\lambda_{\rm obs}}
      {\lambda_{\alpha}}\right)^{3.46} \right] \,d\lambda_{\rm obs}, 
    & \mbox{$1026 < \lambda_{\rm rest} < 1216$ \AA,} \\
    \frac{1}{\Delta \lambda_{B}} \times & \\[4ex]
    \int\limits_{920\left(1+z\right)}^{1015\left(1+z\right)}
    \exp \left[ - \sum\limits_{j=3,5} A_{j} \left( 
      \frac{\lambda_{\rm obs}}{\lambda_{j}}\right)^{3.46}\right] \, d\lambda_{\rm obs}, &
    \mbox{$912 < \lambda_{\rm rest} < 1026$ \AA,} \\
    0, & \mbox{$\lambda_{\rm rest} < 912$ \AA.} \\
  \end{array}
  \right.
  \label{eq:dfac}
\end{equation}
\normalsize
The second line of Equation \ref{eq:dfac} states the contribution from
Ly$\alpha$,  whilst  the  third   line  totals  that  from  Ly$\beta$,
Ly$\gamma$  and Ly$\delta$.   $\Delta \lambda_{A}  = 120\left(  1  + z
\right)$  \AA, $\Delta \lambda_{B}  = 95  \left( 1  + z  \right)$ \AA,
$A_{2}  =  3.6  \times  10^{-3}$  is  the  coefficient  of  Ly$\alpha$
absorption relative to  the Ly$\alpha$ forest contribution. $A_{3}=1.7
\times 10^{-3}$, $A_{4} = 1.2  \times 10^{-3}$ and $A_{5} = 9.3 \times
10^{-4}$   are  the   coefficients  for   Ly$\beta$,   Ly$\gamma$  and
Ly$\delta$, respectively.\par

To determine which filters require the application of $D\left( \lambda
\right)$,  we calculate  $R_{\rm IGM}$  using  Equation \ref{eq:RIGM}:

\begin{equation}
  R_{{\rm IGM}}=\frac{\int\limits_{\lambda_{1}}^{\lambda_{2}}\lambda F\left( \lambda \right) A_{\lambda} D\left( \lambda \right) T \left( \lambda \right) \, d\lambda}{\int\limits_{\lambda_{1}}^{\lambda_{2}}\lambda F\left( \lambda \right) A_{\lambda} T \left( \lambda \right) \, d\lambda}
\label{eq:RIGM}
\end{equation}

This ratio  considers how many  source counts pass through  the filter
both with  and without  IGM absorption, whilst  weighting the  flux at
each  wavelength by  the filter  transmission curve,  $T\left( \lambda
\right)$. A  value of $R_{\rm IGM}=1$  indicates that there  is no IGM
absorption  in the band,  and so  the flux  evolves slowly  across the
bandpass in question. Similarly,  when $R_{\rm IGM}=0$ all flux within
the  filter is  absorbed  by  neutral hydrogen  and  is therefore  not
evolving across the band.  In these instances we can neglect the shape
of the filter transmission curves, and therefore use the first line of
Equation \ref{eq:finflux}.\par

\begin{equation}
  F_{{\rm filter}} = \left\{
  \begin{array}{ll}
     \frac{\int\limits_{\lambda_{1}}^{\lambda_{2}} F \left( \lambda \right)
       A_{\lambda} D \left( \lambda \right)
       \, d\lambda}{\int\limits_{\lambda_{1}}^{\lambda_{2}} \, d\lambda}, &
     \mbox{$R_{{\rm IGM}} = 0,1$,} \\
     R_{IGM}\frac{\int\limits_{\lambda_{1}}^{\lambda_{2}} F \left( \lambda \right)
       A_{\lambda} \, d\lambda}{\int\limits_{\lambda_{1}}^{\lambda_{2}} \, d\lambda}, &
     \mbox{$0<R_{{\rm IGM}}<1$.} \\
  \end{array}
  \right.
  \label{eq:finflux}
\end{equation}

When  the  effects  of   neutral  hydrogen  absorption  lie  within  a
particular filter,  there is a sharp transition  within its wavelength
coverage,  $0 <  R_{\rm IGM}  <  1$, and  the filter  requires a  more
detailed treatment. In such a filter there is a fraction of wavelength
coverage  which  experiences suppression  of  GRB  flux.   We need  to
calculate this fraction so we  can correctly predict the total average
magnitude across  the filter. We  therefore apply a factor  of $R_{\rm
  IGM}$  to the  filter magnitude,  which correctly  accounts  for the
fraction of  flux that successfully  reaches the telescope  and passes
through the filter  in question.  This is shown in  the second line of
Equation \ref{eq:finflux}.\par

\subsection{Fitting and prior probabilities}
\label{sec:priors}

To  implement the  model described  we used  the {\sc  amoeba} fitting
routine within  {\sc idl} as well  as the {\sc  idl} Astronomy Library
\citep{1993ASPC...52..246L}. {\sc amoeba} is a robust routine that can
avoid local minima in parameter space.\par

To avoid  regions of the parameter space  containing unphysical values
of $\beta_{\rm opt}$ we  imposed a prior probability distribution upon
the spectral index in the optical and NIR regime. This was implemented
in the manner described in \S 2 of \citet{2001ApJ...553..235R}. As the
afterglow  emission   is  attributed  to   synchrotron  radiation,  we
considered three possible regimes: (i)  the optical regime lies on the
same   power-law   segment   as   the   X-ray   regime   ($\beta_{{\rm
    opt}}=\beta_{{\rm X}}$), (ii) there is a cooling break between the
two regimes in the fast  cooling regime (implying $\beta_{{\rm opt}} =
\frac{1}{2}$),  (iii)  a cooling  break  in  the  slow cooling  regime
($\beta_{{\rm            opt}}=\beta_{{\rm           X}}-\frac{1}{2}$)
\citep{2002ApJ...568..820G}.   The  cooling   break   occurs  at   the
synchrotron frequency  emitted by electron with a  Lorentz factor that
causes the cooling timescale  from radiation, $t_{\rm cool}$, to equal
the    dynamical     time    of    the     system,    $t_{\rm    dyn}$
\citep{2000ApJ...534L.163G}.  In  the   fast  cooling  regime  $t_{\rm
  cool}<t_{\rm dyn}$  and electrons  cool via radiative  losses faster
than the dynamical timescale. The  slow cooling regime is the reverse,
in which the  majority of electrons are unable  to cool within $t_{\rm
  dyn}$.\par

These three regimes for $\beta_{\rm  opt}$ lead to a prior probability
$p  \left( \beta_{\rm opt}  | I  \right)$, given  in full  in Equation
\ref{eq:beta_opt}.  $I$  is the known information  about the parameter
being fitted (in our prior this is $\beta_{\rm opt}$).\par
 
\footnotesize
\begin{equation}
\begin{array}{ll}
  p \left( \beta_{\rm opt} | I \right) = & \frac{1}{2}G\left( \beta_{\rm opt}, \beta_{\rm X}, \Delta \beta_{\rm X} \right) +  \frac{1}{4}G\left( \beta_{\rm opt}, \frac{1}{2},\Delta\beta_{\rm X}\right) +\\[1ex]
    & \frac{1}{4}G\left( \beta_{\rm opt},\beta_{\rm X} - \frac{1}{2},\Delta\beta_{\rm X}\right) .\\
\end{array}
  \label{eq:beta_opt}
\end{equation}
\normalsize
\citet{2012A&A...537A..15S}   discovered  that   approximately  50$\%$
(25/49) of their sample of X-ray and NIR detected long GRB SEDs can be
well fitted  assuming $\beta_{\rm opt}=\beta_{\rm X}$,  so we weighted
$p \left(  \beta_{\rm opt} |  I \right)$ accordingly. With  no further
information  on whether  the  burst is  in  the fast  or slow  cooling
regime,  these two options  were equally  weighted to  be 25\%  of the
total distribution each.\par

In this instance our prior is the three expected values of $\beta_{\rm
  opt}$ already discussed.  Effectively, the prior probability weights
the parameter space and reduces the total viable range a parameter can
take.  We include realistic estimates of the Gaussian width, $\sigma$,
for  each  of the  distributions  in $p  \left(  \beta_{\rm  opt} |  I
\right)$ and  so ensure  $\beta_{\rm opt}$ is  not forced to  take the
exact value of one of three states discussed.\par

\subsection{Error estimation in $z_{\rm phot}$}
\label{sec:zerr}

The primary  output of  the template fitting  routines is a  value for
$z_{\rm  phot}$.  To  obtain  an error  in  this value  we consider  a
uniform grid of values for redshift, $z_{\rm grid}$. This grid extends
over the full range of potential  redshifts that can be fitted (see \S
\ref{sec:tests}). At  each point of $z_{\rm grid}$  we fit $\beta_{\rm
  opt}$, $A_{V}$  (for the no, MW,  LMC and SMC type  dust models) and
the  normalization  to the  intrinsic  spectrum,  $N_{0}$, using  {\sc
  amoeba}  whilst holding  $z_{\rm grid}$  constant.  To  optimize the
search of the  parameter space using {\sc amoeba}  we invoke $p \left(
\beta_{\rm  opt}  | I  \right)$.   To do  so  we  calculate $p  \left(
\beta_{\rm opt} | I \right)$ within the fitting routine and include it
as an additional  term of our fit statistic,  $\chi^{2}_{\rm eff}$, as
shown  in  Equation  \ref{eq:chisqeff}.  The first  term  of  Equation
\ref{eq:chisqeff} corresponds to the standard $\chi^{2}$ value.\par

\begin{equation}
  \chi^{2}_{\rm eff} = \chi^{2} -2 \log \left( p \left( \beta_{\rm opt} | I \right) \right).
\label{eq:chisqeff}
\end{equation}

By calculating $\chi^{2}_{\rm eff}$ we  force {\sc amoeba} to find the
best fit solution  at each point of $z_{\rm  grid}$ which accounts for
our knowledge  of $\beta_{\rm X}$. We  can then find a  measure of the
prior   weighted  probability  distribution,   $p\left(  D   |  \theta
I\right)$, where  $\theta$ corresponds to the full  set of parameters,
directly  from $\chi^{2}_{\rm eff}$  at each  value of  $z_{\rm grid}$
using Equation  \ref{eq:probdist}.  The value of  $z_{\rm grid}$ where
$p\left(D  |  \theta  I\right)$  peaks  is  our  estimate  of  $z_{\rm
  phot}$.\par

\begin{equation}
  p\left( D |  \theta I \right) \propto \exp\left(-\frac{\chi^{2}_{\rm eff}}{2}\right) .
  \label{eq:probdist}
\end{equation}

To  provide an  error  estimate on  $z_{\rm  phot}$ we  then find  the
narrowest range of  $z_{\rm grid}$ containing 90\% and  99.73\% of the
total  weighted probability distribution.   The former  corresponds to
the 90\%  confidence interval, whilst  the latter would  correspond to
the  3$\sigma$ confidence  interval of  a Gaussian  distribution. Both
were  obtained  for  all   four  extinction  models  discussed  in  \S
\ref{sec:dustabs}.\par

\section{Observations of GRB~130606A}
\label{sec:obs}

The  Burst  Alert  Telescope (BAT;  \citealt{2005SSRv..120..143B})  on
board \textit{Swift} triggered  on GRB~130606A (\textit{Swift} trigger
557589)    at   21:04:39    UT   on    2013   June    6$^{{\rm   th}}$
\citep{2013GCN..14781...1U}.  BAT  identified prompt structure  at the
trigger   time   and   later,   brighter  $\gamma$-ray   emission   at
approximately 150 seconds after the initial trigger time.\par

\textit{Swift}/XRT detected  an   uncatalogued  fading   X-ray  source,
providing  a more  accurate position  for ground-based  follow-up with
narrow field instruments.\par

The  enhanced-XRT  position for  GRB~130606A  was  found  to be  ${\rm
  RA(J2000)} =  16^{{\rm h}}37^{{\rm m}}35\fs12$,  ${\rm Dec(J2000)} =
+29\degr  47\arcmin46\farcs4$   with  an  uncertainty   of  $1\farcs5$
\citep{2013GCN..14811...1O}.   This  uncertainty  is  a  radius  which
signifies  90\% confidence.   This position  was calculated  using the
\textit{Swift}/UVOT  to  astrometrically   correct  the  positions  of
sources in the XRT field of view.\par

\citet{2013ApJ...774...26C} observed the afterglow of GRB~130606A with
both the Blue  Channel spectrograph \citep{1989PASP..101..713S} on the
Multiple   Mirror  Telescope   (MMT)  and   the   Gemini  Multi-Object
Spectrograph (GMOS; \citealt{2004PASP..116..425H}) on the Gemini North
telescope.  These  observations had midpoints  of 7.68 and  13.1 hours
after the initial  BAT trigger, for the Blue  Channel spectrograph and
Gemini-N/GMOS  respectively.   From this  rapid  follow-up with  large
aperture facilities,  \citet{2013ApJ...774...26C} obtained high signal
to noise ratio spectra allowing them to measure $z=5.913$.\par

\begin{figure*}
  \begin{center}
    \includegraphics[width=6.5cm,angle=0]{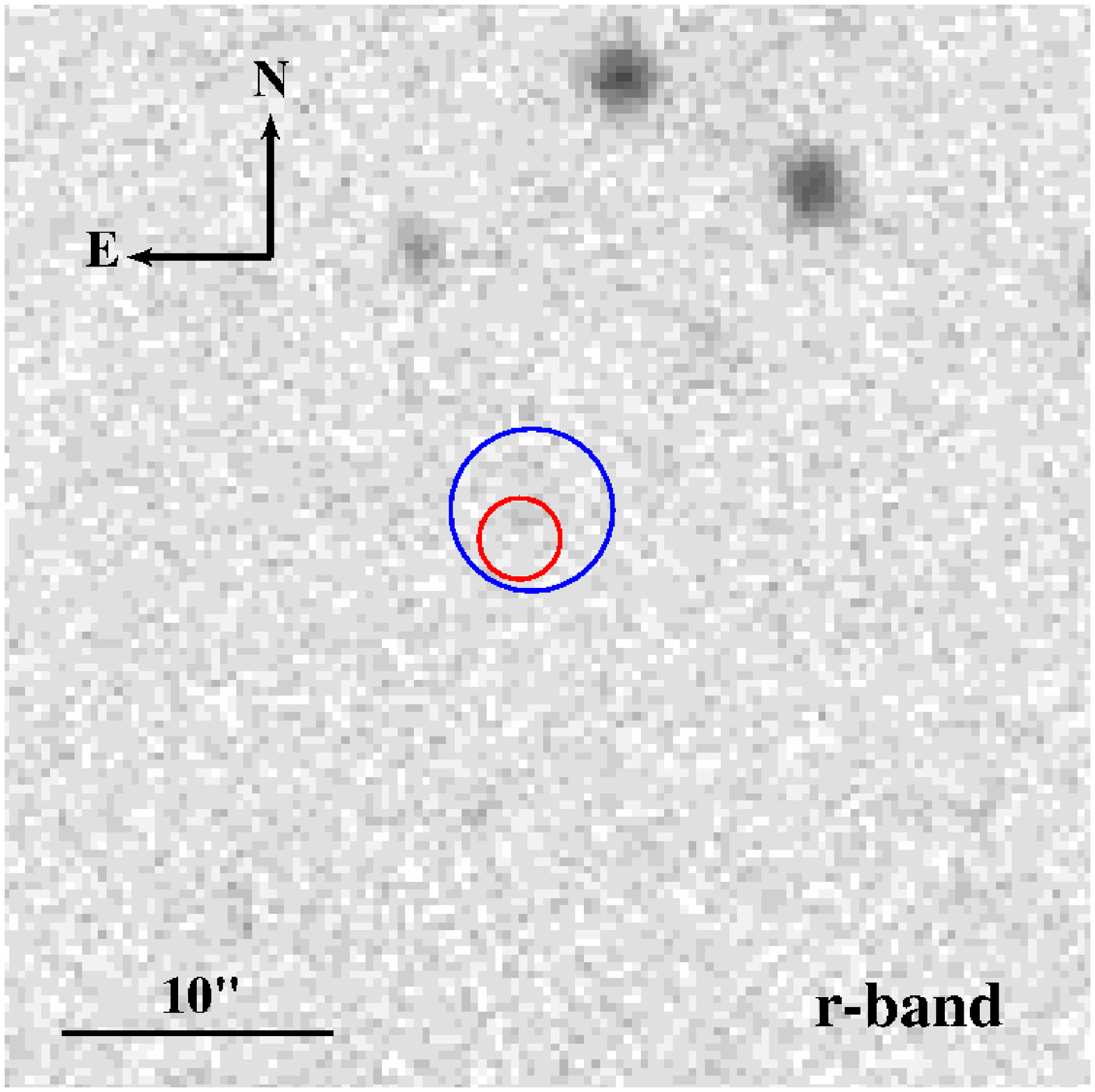}\quad
    \includegraphics[width=6.5cm,angle=0]{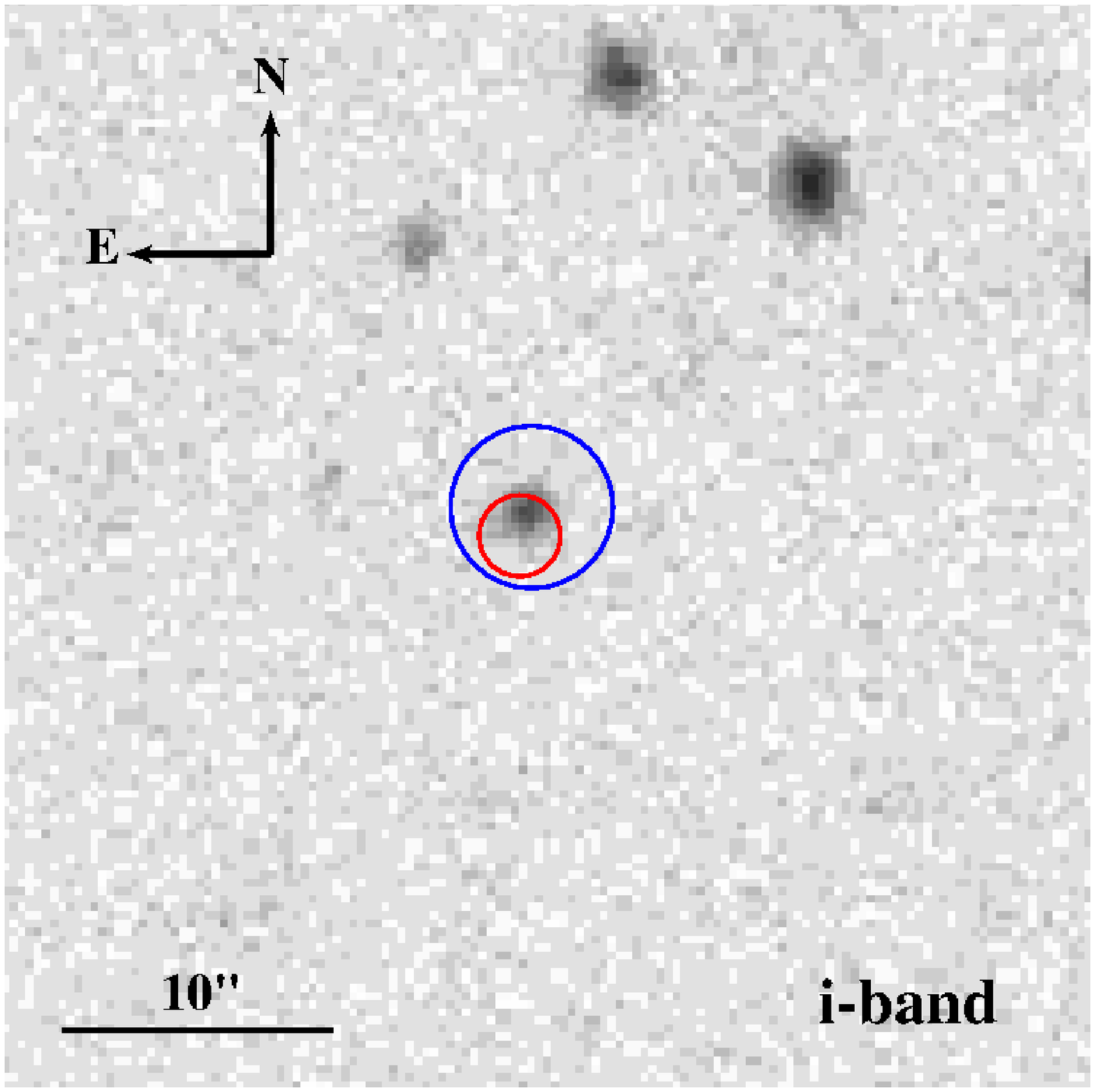}\\[2ex]
    \includegraphics[width=6.5cm,angle=0]{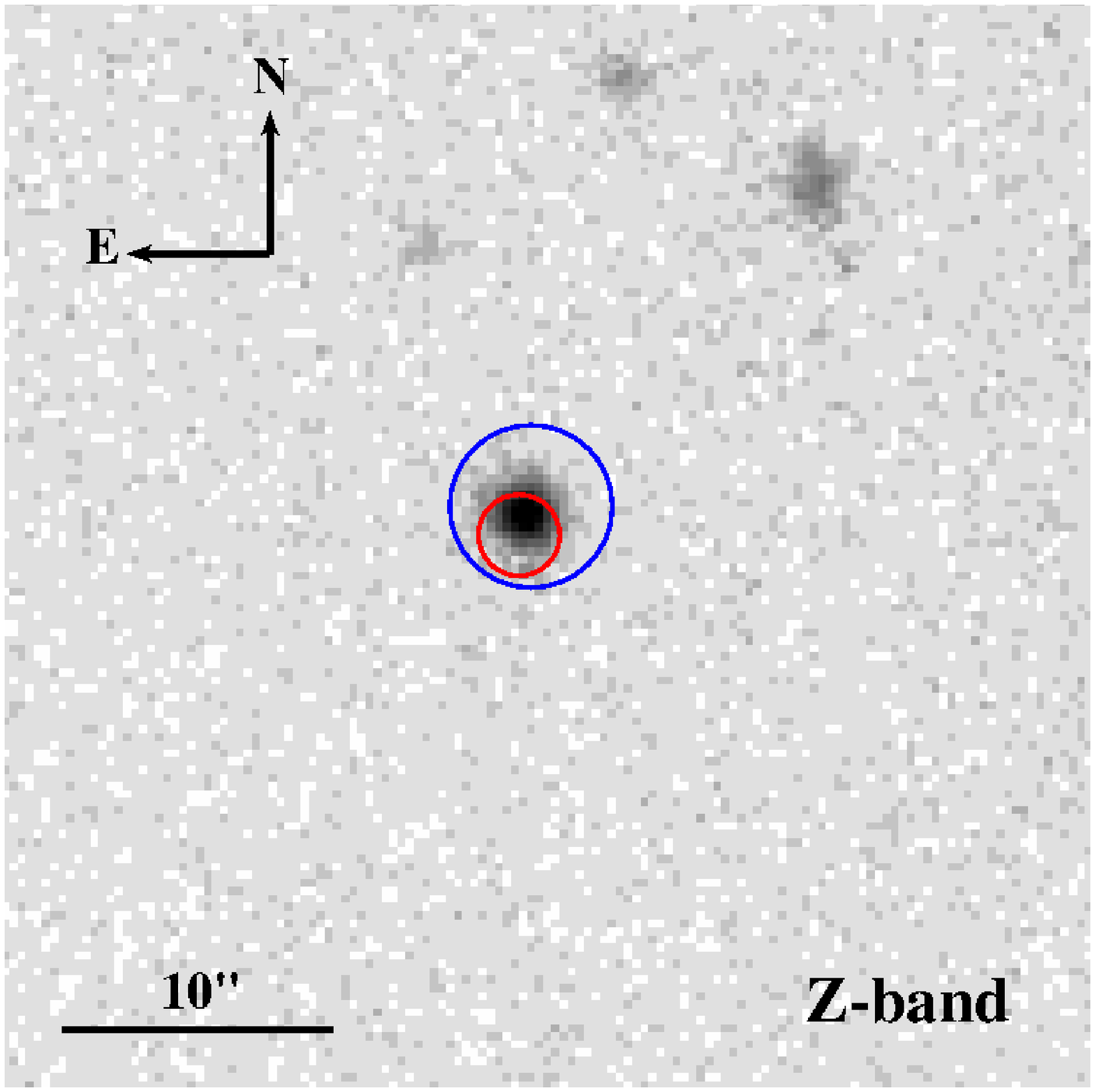}\quad
    \includegraphics[width=6.5cm,angle=0]{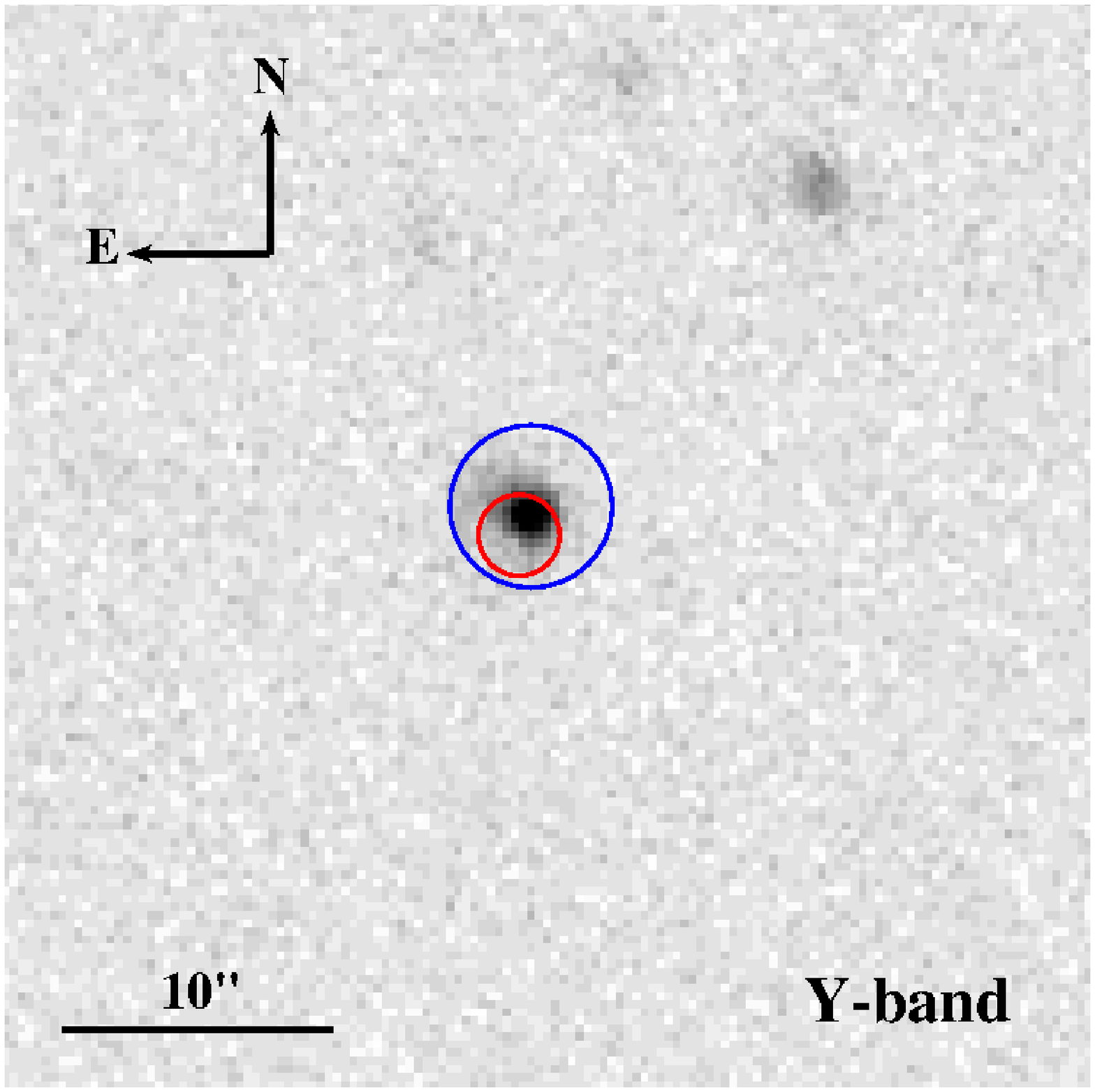}\\[2ex]
    \includegraphics[width=6.5cm,angle=0]{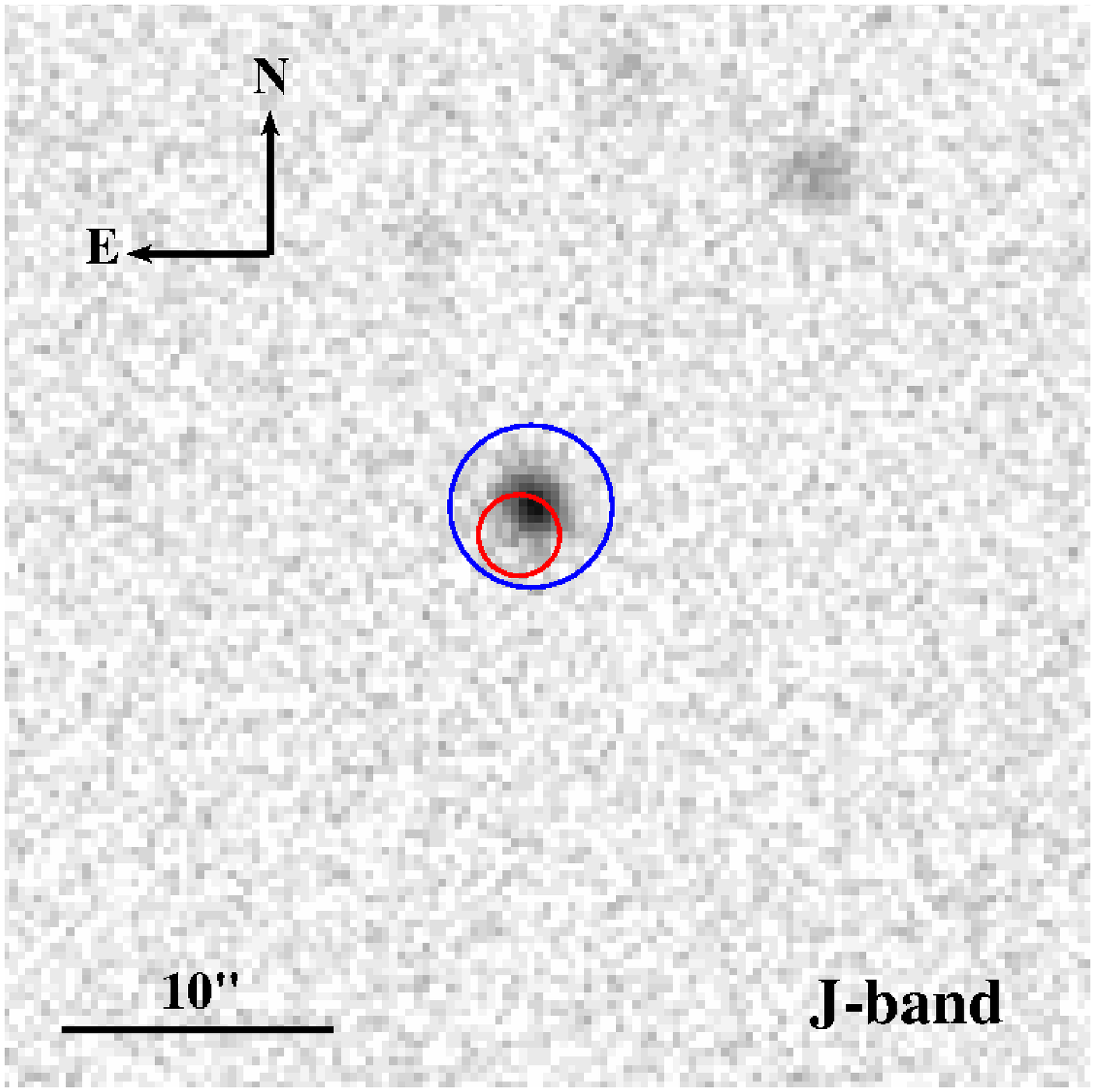}\quad
    \includegraphics[width=6.5cm,angle=0]{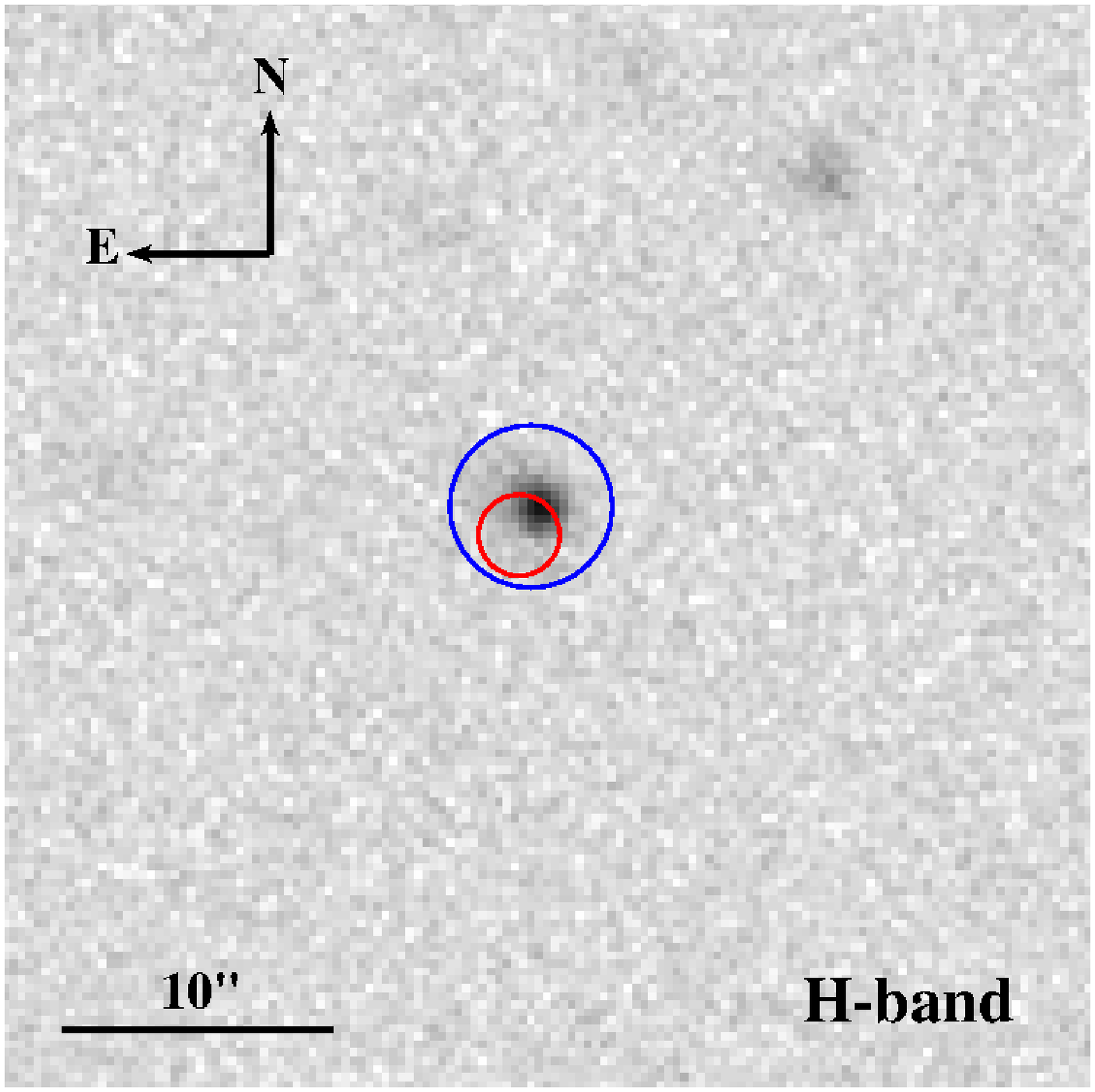}
  \end{center}
  \caption{Photometry obtained for the field of GRB~130606A. Top left:
    \textit{r}-band,   top   right:   \textit{i}-band,  middle   left:
    \textit{Z}-band,  middle   right:  \textit{Y}-band,  bottom  left:
    \textit{J}-band and bottom  right: \textit{H}-band.  Each image is
    the  sum   of  all  observations   taken  on  the  night   of  the
    \textit{Swift}/BAT trigger.   The red circle in each  panel is the
    refined  \textit{Swift}/XRT  error  circle  and  the  blue  circle
    indicates  the  RATIR  identified  counterpart.   All  images  are
    approximately 40$^{\prime\prime}$ $\times$ 40$^{\prime\prime}$ in scale.}
  \label{fig:obs_frames}
\end{figure*}

RATIR first observed  the field of GRB~130606A between  7.38 and 14.19
hours after  the BAT trigger.  Earlier observations were  precluded by
the   GRB  occurring   during  daylight   hours  at   the  Observatorio
Astron\'{o}mico Nacional. RATIR obtained 4.42 hours of exposure in the
\textit{r}- and  \textit{i}-bands and  1.85 hours in  the \textit{Z}-,
\textit{Y}-,          \textit{J}-         and         \textit{H}-bands
\citep{2013GCN..14799...1B}. The observations  for all six photometric
bands    from    this    first    night   are    shown    in    Figure
\ref{fig:obs_frames}.\par

A second epoch of observations  were taken the following night between
31.14   and    37.86   hours   after    the   \textit{Swift}   trigger
\citep{2013GCN..14824...1B}. With  4.96 hours of exposure  time in the
\textit{r}- and  \textit{i}-bands both yielding  upper limits, showing
that the source had faded  in the \textit{i}-band.  The burst remained
bright enough  in each of the  NIR filters to allow  detections in all
four  bands with  a total  exposure time  of 2.07  hours. In  all four
filters the burst  had faded by more than two  magnitudes, as shown in
Table \ref{tab:RATIR_obs}.\par

The  RATIR camera  consists of  two optical  and two  infrared cameras
\citep{2012SPIE.8446E..10B},  allowing for simultaneous  image capture
in four bands ($riZJ$ or $riYH$).  Split filters immediately above the
infra-red detectors allow for near-simultaneous images in six bands by
dithering the  source across the  infra-red detector focal  plane.  We
employ a  dither pattern that  allows for sequential exposure  in $ZJ$
followed by  $YH$.  Dithering also  allows for subtraction of  the sky
background and  the detector dark  signal at the source  position.  We
capture 80s exposure frames in  $ri$ and 67s exposure frames in $ZYJH$
due  to  additional  overhead.   We  apply the  same  image  reduction
pipeline -- with twilight  flat division and bias subtraction routines
written in python and using astrometry.net \citep{2010AJ....139.1782L}
for  image alignment and  {\sc swarp}  \citep{2010ascl.soft10068B} for
image co-addition -- to data taken from each camera.\par

Photometry    is    calculated    by    running    {\sc    sextractor}
\citep{1996A&AS..117..393B} over individual science frames and mosaics
over  a  range of  apertures  between 2  and  30  pixels diameter.   A
weighted average of the flux in this set of apertures for all stars in
a   given    field   is   then   used   to    construct   an   annular
point-spread-function (PSF).   This PSF is then fitted  to the annular
flux  values for  each source  to optimize  signal-to-noise  for point
source photometry.  We perform a field-photometric calibration for the
$riZ$ bands  using the Sloan Digital  Sky Survey Date  Release 9 (SDSS
DR9;\citealt{2012ApJS..203...21A}).  The RATIR $riZ$ filters have been
shown to be  equivalent to the SDSS filters  ($riz$) at the $\lesssim$
3\%  level  \citep{2013Butlerprep}.   We  calibrate  the  $YJH$  bands
relative    to   the    Two    Micron   All    Sky   Survey    (2MASS;
\citealt{2006AJ....131.1163S}), employing  the United Kingdom Infrared
Telescope       (UKIRT)      Wide      Field       Camera      (WFCAM;
\citealt{2009MNRAS.394..675H,2007A&A...467..777C})   to  estimate  $Y$
from $JH$ for field stars.   Photometric residuals for these bands are
also approximately 3\%.\par

\begin{deluxetable*}{cccccccc}
\tablewidth{0pt}
\tabletypesize{\scriptsize}
\tablecaption{RATIR  photometry for  GRB~130606A. The  first  two columns
    denote  the start  and  stop times  ($T_{\rm  start}$ and  $T_{\rm
      stop}$ respectively) of the observations on both nights in hours
    since the BAT trigger. The magnitudes for each filter are given in
    the AB magnitude system.\label{tab:RATIR_obs}}
\tablehead{\colhead{$T_{\rm start}$} & \colhead{$T_{\rm stop}$} & \colhead{$r$} & \colhead{$i$} & \colhead{$Z$} & \colhead{$Y$} & \colhead{$J$} & \colhead{$H$} \\
\colhead{(hours)} & \colhead{(hours)} & &  &  &  &  & 
}
\startdata
      7.38 & 14.19 & 24.48 $\pm$0.30 & 21.80 $\pm$0.06 & 19.32$\pm$0.02 & 
      19.11$\pm$0.02 & 18.96 $\pm$0.02 & 18.64 $\pm$0.02 \\
      31.14 & 37.86 & $>$ 24.06 & $>$ 23.95 & 21.50$\pm$0.09 &
      21.41$\pm$0.12 & 21.16$\pm$0.12 & 20.78$\pm$0.12 \\
\enddata
\vspace{-0.25cm}
\end{deluxetable*}

\section{Modeling GRB~130606A SED}
\label{sec:130606A}

The spectroscopic observations for  GRB~130606A allow us to assess the
accuracy of  our template fitting routine  using this burst  as a test
case.  The  standard operation of  our algorithm requires  input RATIR
photometry and an  estimation of $\beta_{\rm X}$. The  former is taken
from the initial RATIR  GRB Coordinates Network (GCN) circular, whilst
the latter is obtained from  the UK \textit{Swift} Science Data Centre
(UKSSDC)              automated              analysis             data
products\footnote{http://www.swift.ac.uk/xrt\_spectra/}
\citep{2009MNRAS.397.1177E}.\par

The algorithm  then uses  a broad 1-dimensional  grid in  redshift ($0
\leqslant z_{\rm  grid} \leqslant 12$)  to find $z_{\rm  phot}$.  This
value provides a rapid and robust estimate of $z_{\rm phot}$ that aids
in  determining whether  follow-up with  large aperture  facilities is
warranted.\par

We  then use  2-dimensional grids  to refine  our estimate  of $z_{\rm
  phot}$. Two such  grids are used for each  extinction law. The first
fits  $A_{V}$ and  model normalization,  $N_{0}$, at  fixed  points in
$z_{\rm  grid}$ $\beta_{\rm opt}$  parameter space  over the  range $0
\leqslant  z_{\rm grid} \leqslant  12$. By  determining the  region of
this parameter space with  the smallest total $\chi^{2}_{\rm eff}$, we
then produce a second grid which focuses on the best fit solution with
a   higher  resolution   in  both   $z_{\rm  grid}$   and  $\beta_{\rm
  opt}$.\par

In practice, the results from  the finer, 2-dimensional grid provide a
more precise  estimate of $z_{\rm phot}$.   However, when coordinating
follow-up  from large  aperture  facilities, those  produced from  the
1-dimensional   grid  $z_{\rm   grid}$  provide   sufficiently  robust
estimates  such that  further  observations can  be  requested at  the
earliest   possible  epoch   after   the  initial   \textit{Swift}/BAT
trigger.\par

For  GRB~130606A we used  RATIR photometry  obtained between  7.38 and
7.79    hours    after    the    \textit{Swift}/BAT    trigger    time
\citep{2013GCN..14799...1B}.   As the  RATIR image  reduction pipeline
runs on  data as it is  available from the instrument,  it was already
possible at  this epoch to identify GRB~130606A  as an \textit{r}-band
dropout      candidate.      We     also      obtained     $\beta_{\rm
  X}=0.86_{-0.13}^{+0.14}$ from the UKSSDC.\par

\begin{figure*}
  \begin{center}
    \includegraphics[width=15cm,clip,angle=0]{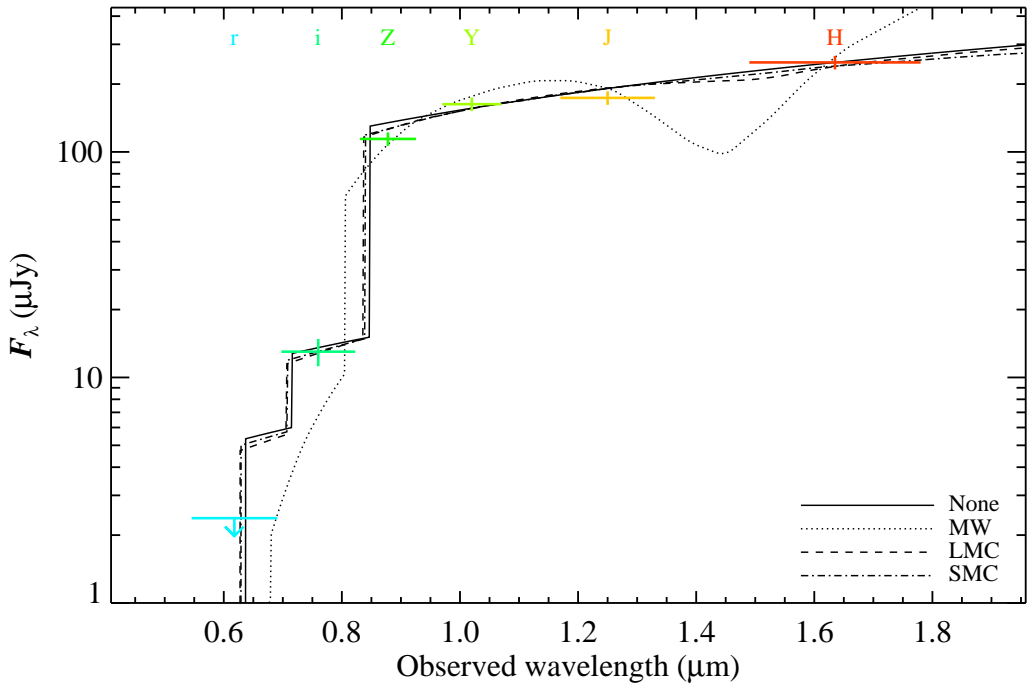}
  \end{center}
  \caption{SED   templates   for   each   of   the   fits   in   Table
    \ref{tab:best_fits}.   The   colored  points  correspond   to  the
    measured RATIR photometry, with the filter being marked above each
    measurement.   The black  lines  indicate the  best fits  obtained
    using the specific host extinction laws detailed in the key.}
  \label{fig:seds}
\end{figure*}

The best fit for each of  the four extinction laws, the three standard
templates from \citet{1992ApJ...395..130P} and  a fit where we assumed
no host  extinction, are  detailed in Table  \ref{tab:best_fits}.  The
corresponding  templates  are plotted  with  the  RATIR photometry  in
Figure \ref{fig:seds}.  It  is important to note that  we only include
dust  extinction from  the host  galaxy. For  GRB~130606A, there  is a
possibility  of  an intervening  system,  with  its  own unknown  dust
content  \citep{2013arXiv1312.3934T}. If  present,  $z_{\rm DLA}=5.8$,
meaning that we would be unable to distinguish between the dust in the
host and any  potential dust in the DLA. Even  should the dust content
of the DLA be high, our  value of $z_{\rm phot}$ remains robust due to
the small difference in redshift better then host galaxy and DLA.\par

\begin{deluxetable}{cccccc}
\tablewidth{0pt}
\tabletypesize{\scriptsize}
\tablecaption{Details of the fitted  solutions to the SED of GRB~130606A,
    stating which host extinction  model was used, $z_{\rm phot}$, its
    $3\sigma$  error, $\beta_{\rm  opt}$, $A_{V}$  and  $\chi^{2}$ for
    each  model.   These solutions  were  obtained  using the  initial
    1-dimensional grid of $z_{\rm grid}$ values.\label{tab:best_fits}}

\tablehead{\colhead{Extinction} & \colhead{$z_{\rm phot}$} & \colhead{$\beta_{\rm opt}$} & \colhead{$A_{V}$} & \colhead{$\chi^{2}/ \nu$}
}
\startdata
 None & 5.97$_{-0.08}^{+0.16}$ & 0.99 & 0.00 & 2.62/3 \\
 MW & 5.62$_{-0.12}^{+0.56}$ & 0.41 & 1.48 & 2.71/2 \\
 LMC & 5.87$_{-0.22}^{+0.31}$ & 0.43 & 0.25 & 4.48/2 \\
 SMC & 5.90$_{-0.22}^{+0.29}$ & 0.42 & 0.13 & 3.59/2 \\
\enddata
\vspace{-0.25cm}
\end{deluxetable}

Table  \ref{tab:best_fits}  suggests that  either  the  MW or  no-dust
solution  best  represents  the  data.  Visual  inspection  of  Figure
\ref{fig:seds}  suggests  the LMC  and  SMC  models  are of  at  least
comparable quality,  however, the magnitude compared  to that measured
by  RATIR  is  an  average  across the  wavelength  coverage  of  each
filter. For the  LMC and SMC templates, the  average values across the
$J$- and $H$-bands are more discrepant than those of the MW or no dust
models. When using a MW type  extinction law, a large quantity of dust
is favored,  with $A_{V}=1.48$.  Consequently, $z_{\rm  phot}$ for the
MW template is reduced, as a large quantity of dust contributes to the
strong   suppression  of   intrinsic  GRB   flux  in   the   $r$-  and
$i$-bands.\par

\begin{figure*}
  \begin{center}
    \includegraphics[width=7.5cm,clip,angle=0]{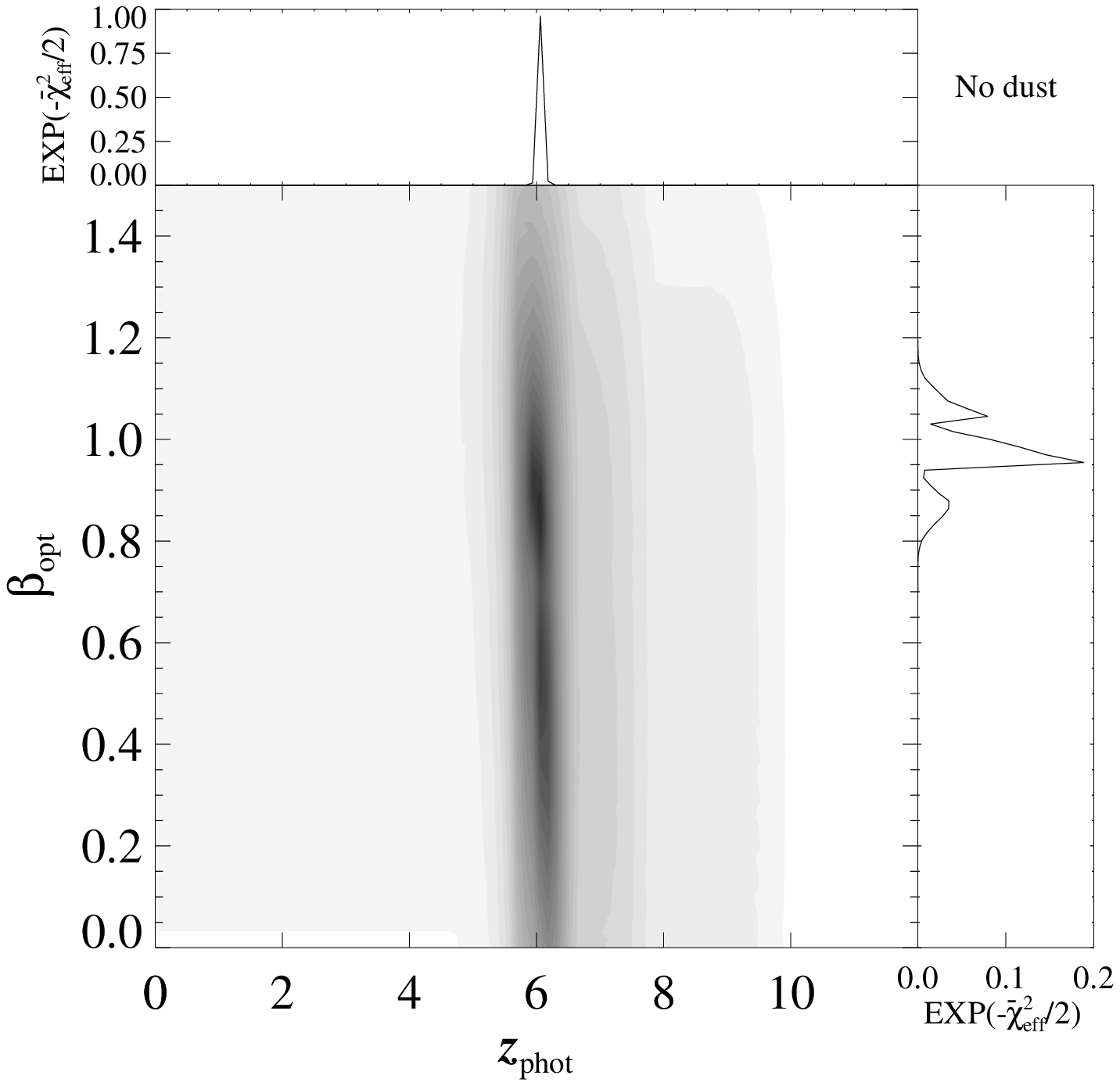}\quad
    \includegraphics[width=7.5cm,clip,angle=0]{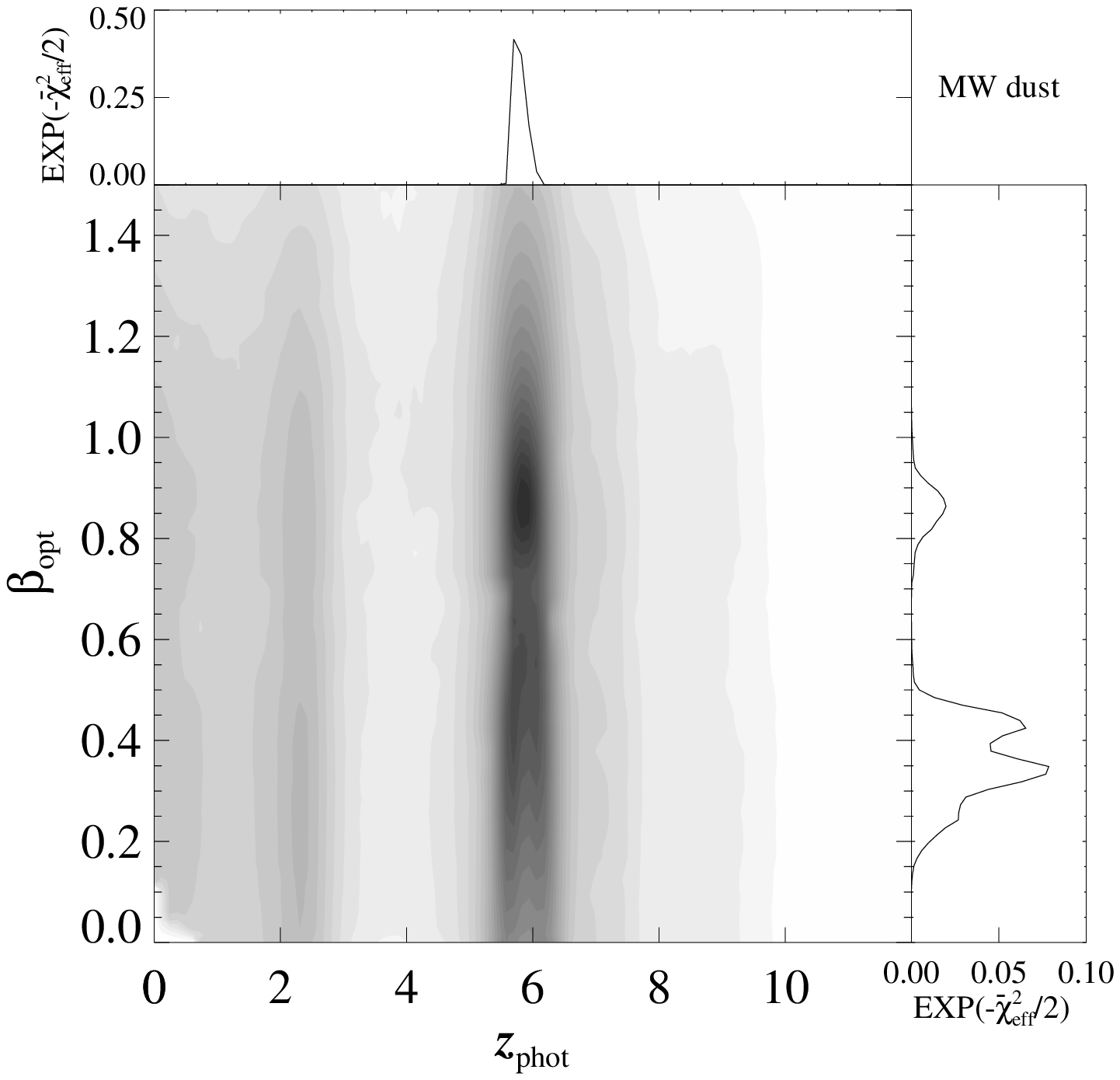}\\
    \includegraphics[width=7.5cm,clip,angle=0]{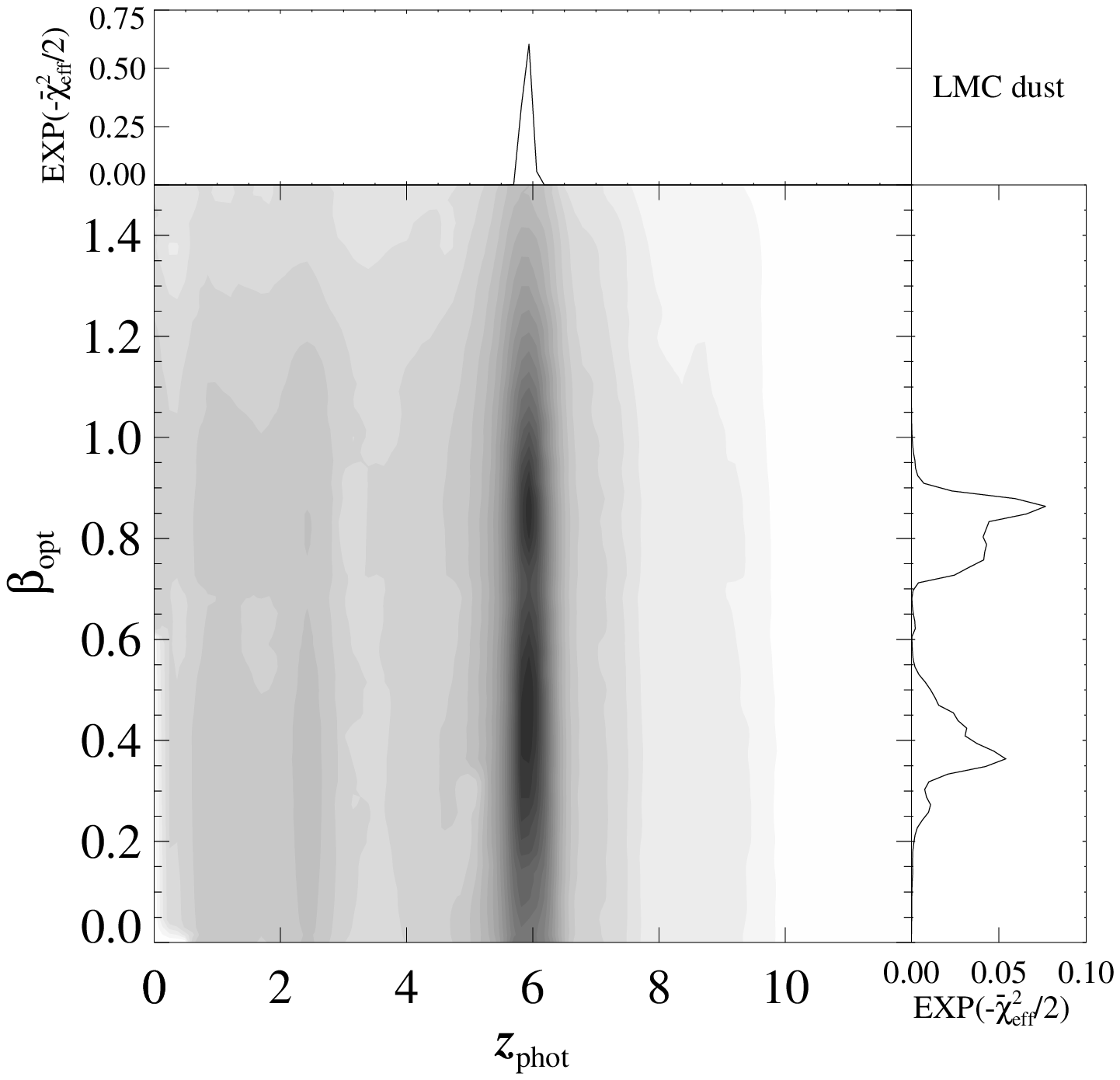}\quad
    \includegraphics[width=7.5cm,clip,angle=0]{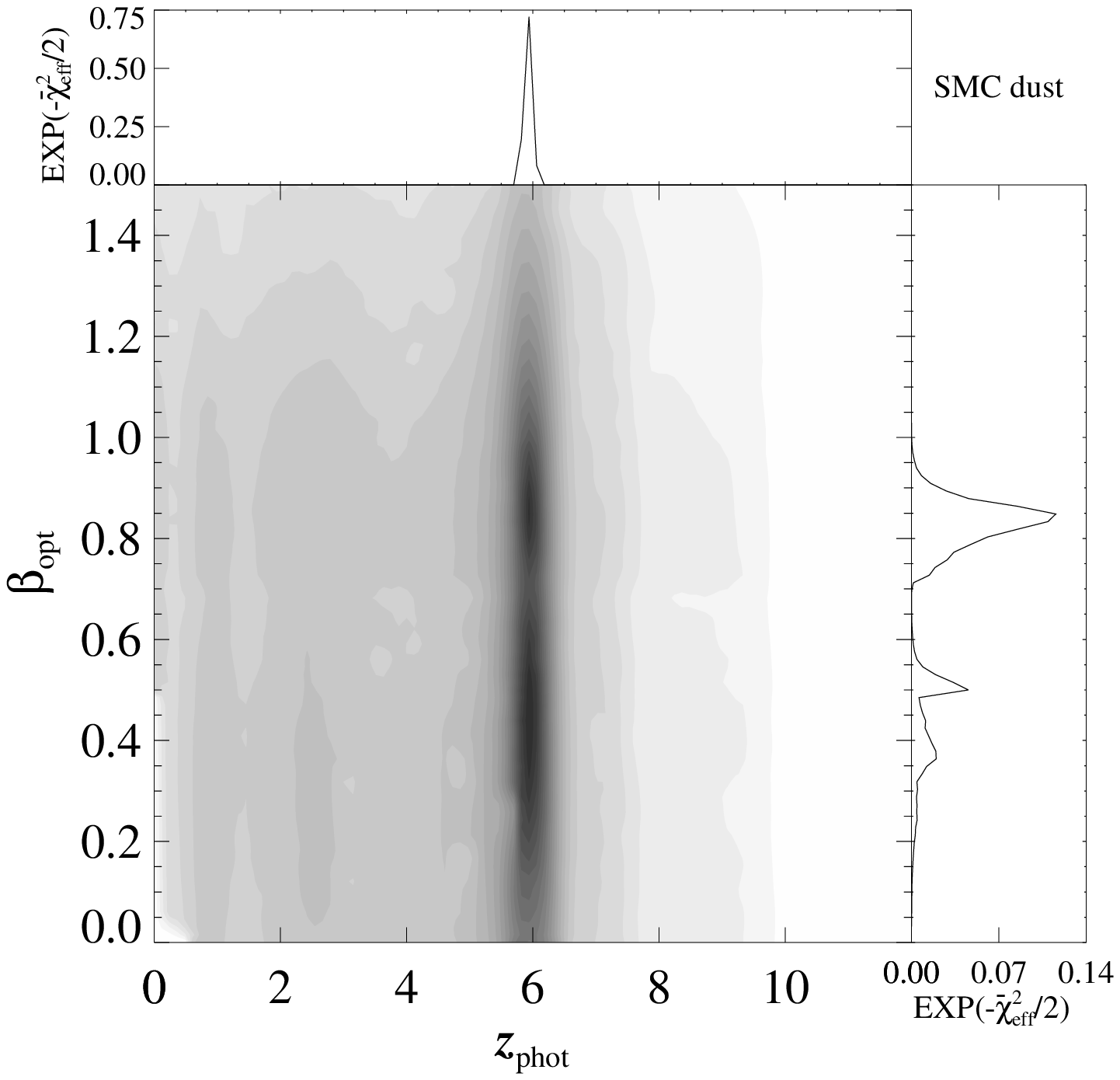}\\
  \end{center}
  \caption{2-dimensional  probability   maps  for  grids   across  the
    parameter   spaces   of   each   solution  presented   in   Figure
    \ref{fig:seds}.   The   horizontal  axis  denotes   the  simulated
    redshift, the  vertical axis denotes the  local intrinsic spectral
    index  of the  GRB afterglow.   The color  scale shows  $\log _{e}
    \chi_{\rm  eff}^{2}$ of  the  fits, where  dark  regions have  the
    lowest  $\chi^{2}$. The narrow  horizontal panels  demonstrate the
    1-dimensional probability distributions for $z_{\rm phot}$, whilst
    the  narrow  vertical panels  show  the 1-dimensional  probability
    distributions of $\beta_{\rm opt}$.  Top left: no host extinction,
    top right:  Milky Way host  extinction law, bottom left:  LMC host
    extinction law, bottom right: SMC host extinction law.}
  \label{fig:conts_big}
\end{figure*}

\begin{figure*}
  \begin{center}
    \includegraphics[width=7.5cm,clip,angle=0]{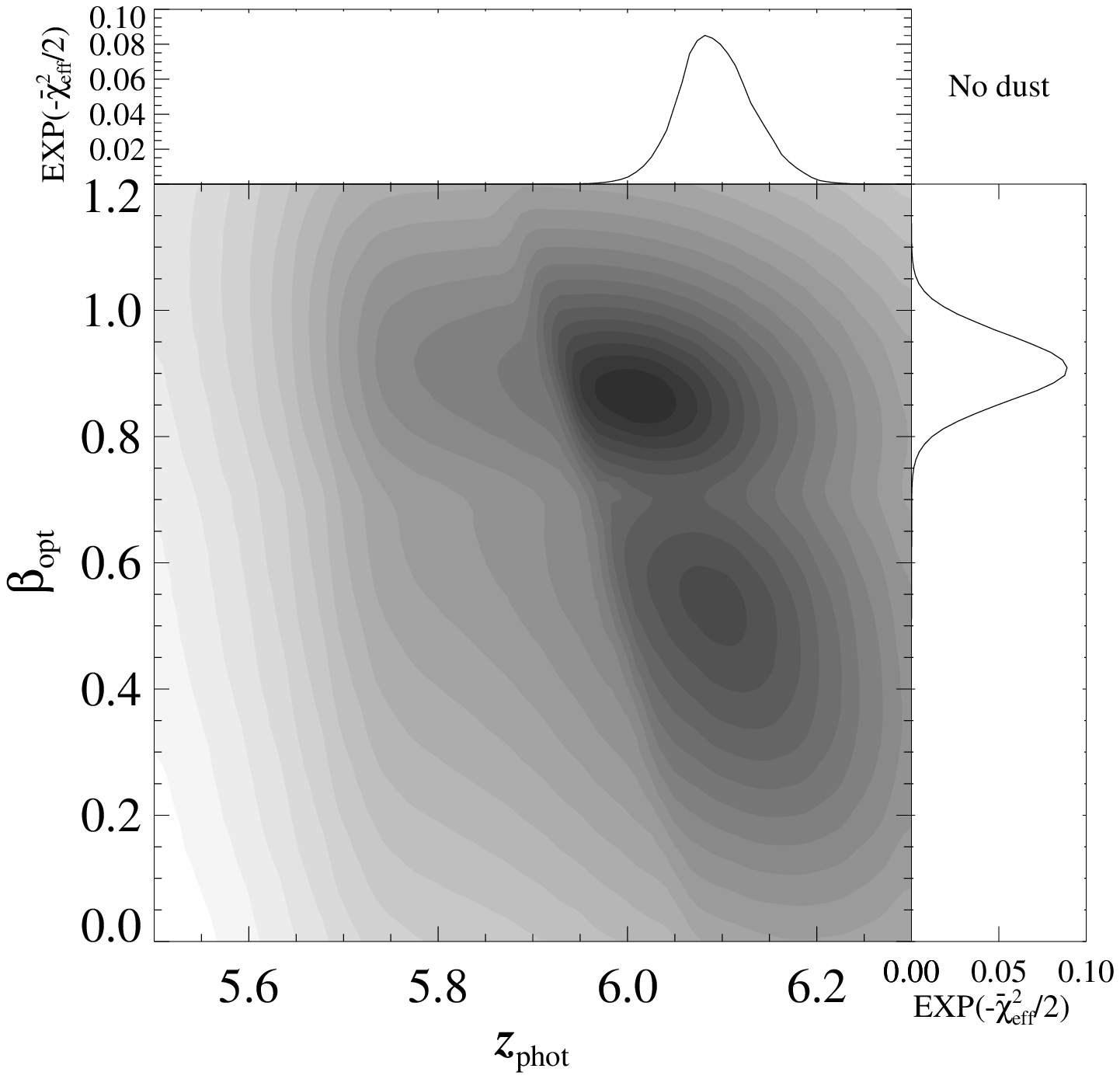}\quad
    \includegraphics[width=7.5cm,clip,angle=0]{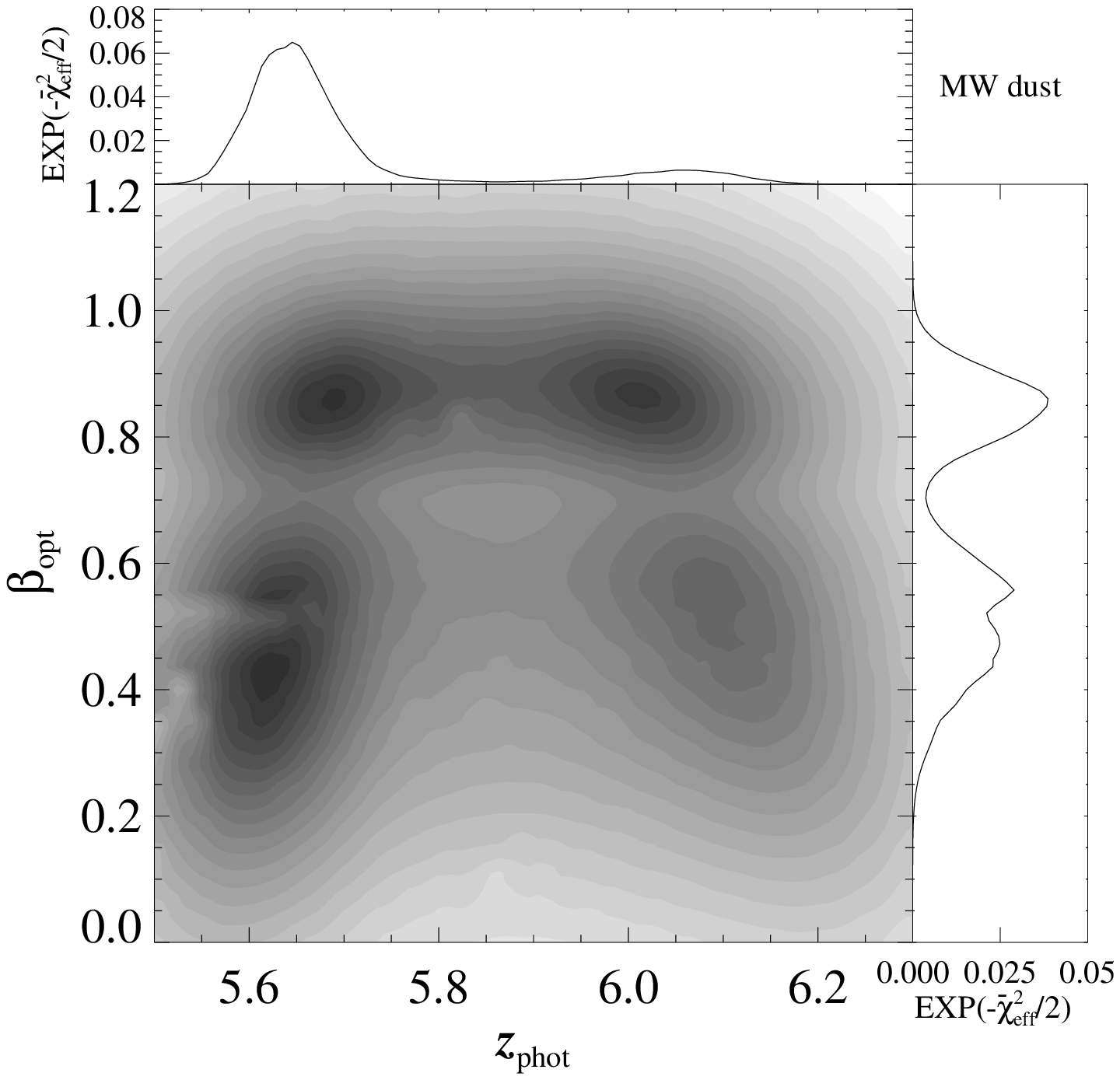}\\
    \includegraphics[width=7.5cm,clip,angle=0]{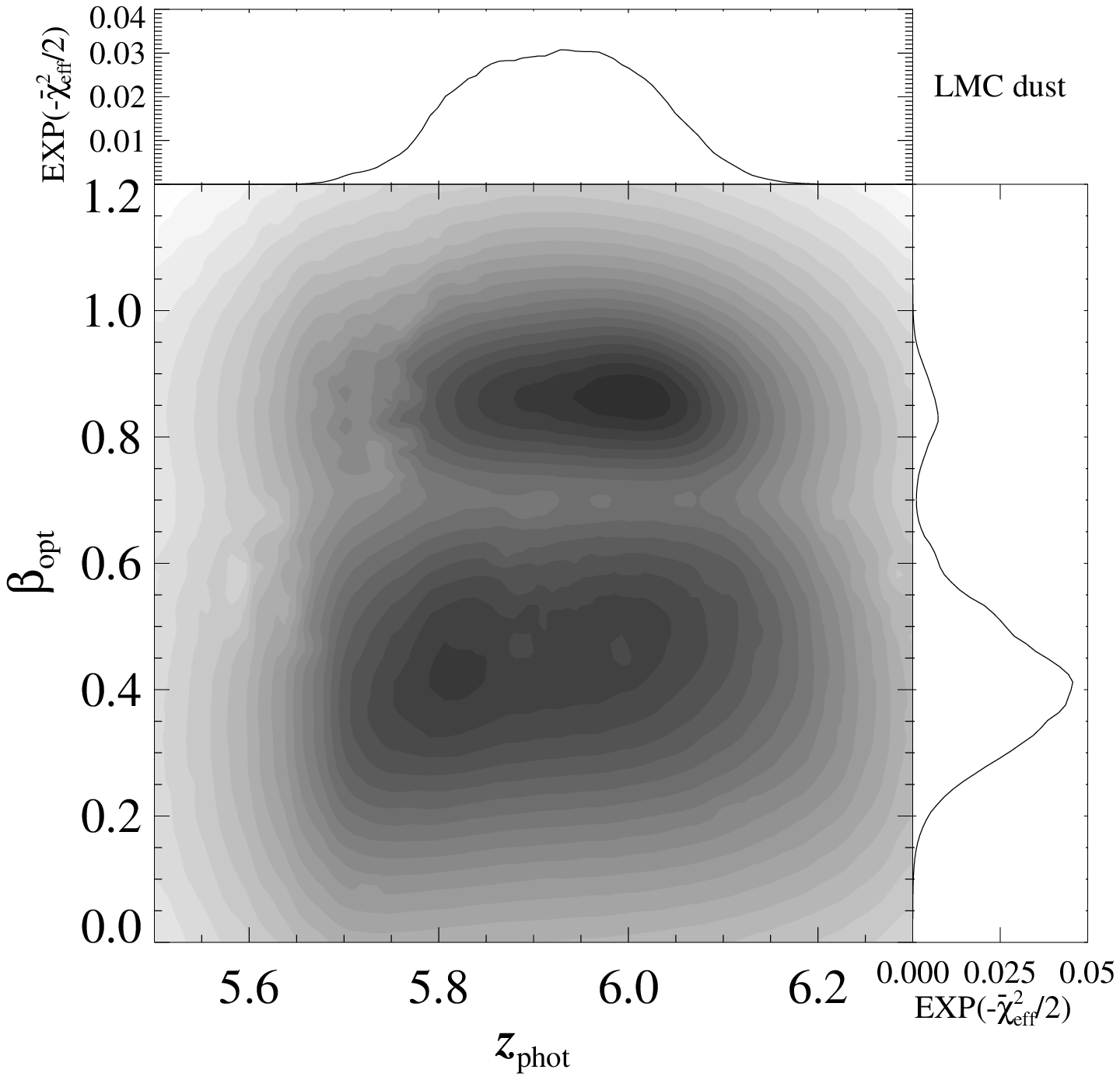}\quad
    \includegraphics[width=7.5cm,clip,angle=0]{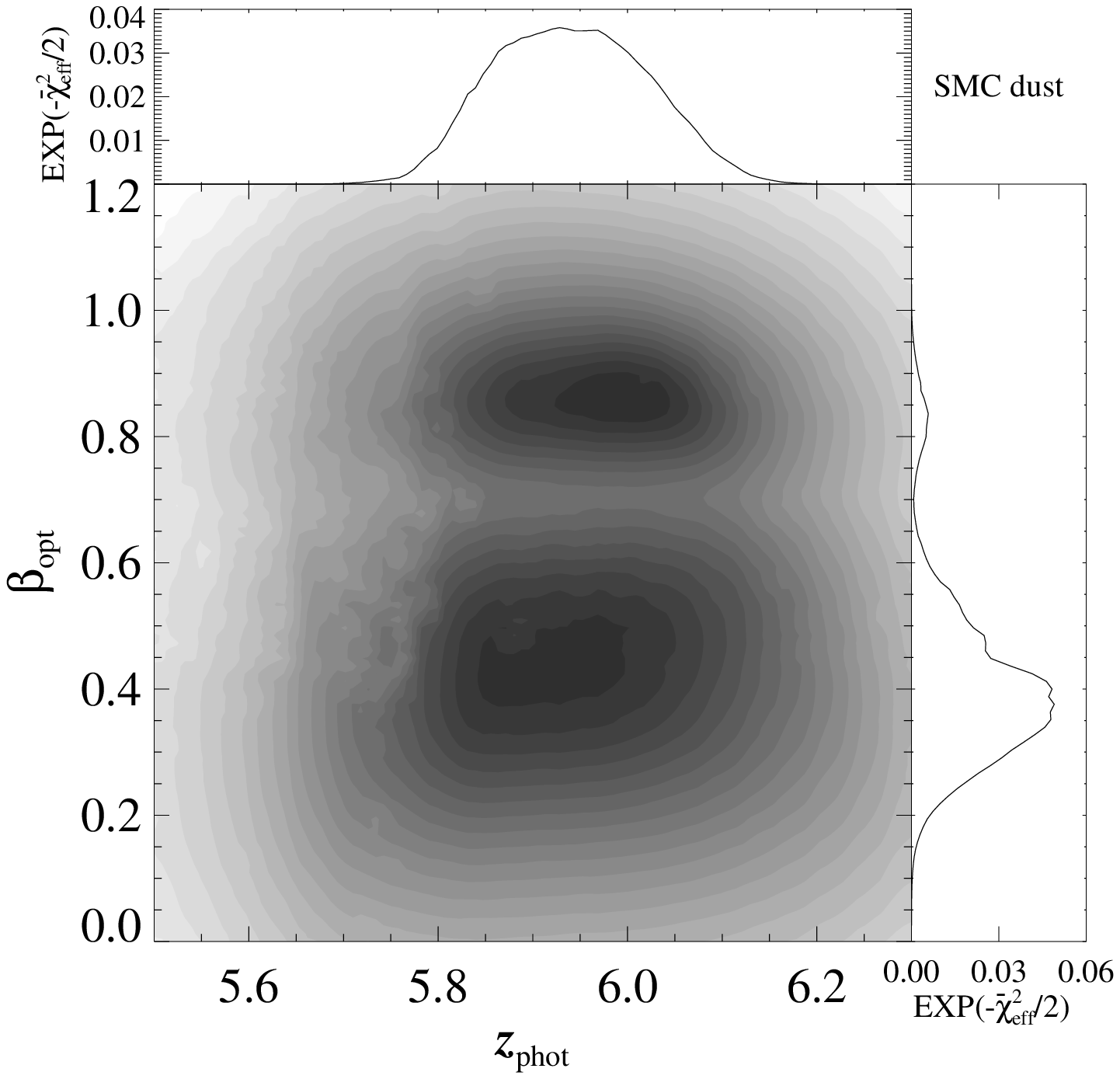}\\
  \end{center}
  \caption{2-dimensional  probability maps  for grids  focused  on the
    region  of  best fit  of  each  parameter  space shown  in  Figure
    \ref{fig:conts_big}.   The horizontal  axis denotes  the simulated
    redshift  and  the  vertical  axis  denotes  the  local  intrinsic
    spectral index of the GRB afterglow.  The color scale of the large
    panels shows $\log _{e} \chi^{2}$  of the fits, where dark regions
    have  the   lowest  $\chi^{2}$.   The   narrow  horizontal  panels
    demonstrate   the  1-dimensional  probability   distributions  for
    $z_{\rm  phot}$,  whilst  the  narrow  vertical  panels  show  the
    1-dimensional probability distributions  of $\beta_{\rm opt}$. Top
    left:  no host extinction,  top right:  Milky Way  host extinction
    law, bottom left: LMC host  extinction law, bottom right: SMC host
    extinction law.}
  \label{fig:conts_small}
\end{figure*}

Figure \ref{fig:conts_big},  which contains the  probability maps over
all allowed ranges of $z_{\rm grid}$, shows the best fit solutions are
located in  a narrow redshift range  at $z_{\rm phot} \sim  6$ for all
four extinction laws. The  finer resolution probability maps in Figure
\ref{fig:conts_small}   allow   more    detailed   structure   to   be
discerned. In the no dust model there are two regions of the parameter
space where the prior  weighted probability is maximized.  The details
of the  local maximum  of each are  in Table  \ref{tab:all_fits}.  The
best  found solution  has  $\beta_{\rm opt}  \sim  \beta_{\rm X}$  and
corresponds to $z_{\rm phot}=6.01$. This is in good agreement with the
cruder estimate  made using the 1-dimensional grid  in redshift, which
is  $z_{\rm  phot}=5.97_{-0.08}^{+0.16}$  for  an  SED  with  no  dust
extinction (see Table \ref{tab:best_fits}).\par

The  LMC  and SMC  probability  maps  in Figure  \ref{fig:conts_small}
contain similar  structure. The first solution is  at $\beta_{\rm opt}
\sim \beta_{\rm X}$, as found  when $A_{V}=0$. With the inclusion of a
non-zero $A_{V}$ this solution is  more extended in the $z_{\rm phot}$
axis. This is  because at marginally lower redshifts  a higher $A_{V}$
allows absorption from the IGM  to contribute to the flux deficit seen
at shorter wavelengths. This elongation  in the $z_{\rm phot}$ axis is
more pronounced in the lower  $\beta_{\rm opt}$ solution seen for both
the LMC and  SMC extinction laws. Some values  of $\beta_{\rm opt}$ do
not exactly coincide with the expected values resulting from a cooling
break.   These result  from an  enhancement in  the  prior probability
distribution  between  $\beta_{\rm  opt}  = \beta_{\rm  X}  -0.5$  and
$\beta_{\rm  opt}  =  0.5$  where  the slow  and  fast  cooling  break
solutions overlap.\par

As  with  the $A_{V}=0$  extinction  law,  the  $\beta_{\rm opt}  \sim
\beta_{\rm X}$  is the  better of the  two alternative  solutions, and
indeed a fitted  value of $A_{V}=0$ is retrieved for  both the LMC and
SMC extinction laws.\par

\begin{figure*}
  \begin{center}
    \includegraphics[width=15cm,clip,angle=0]{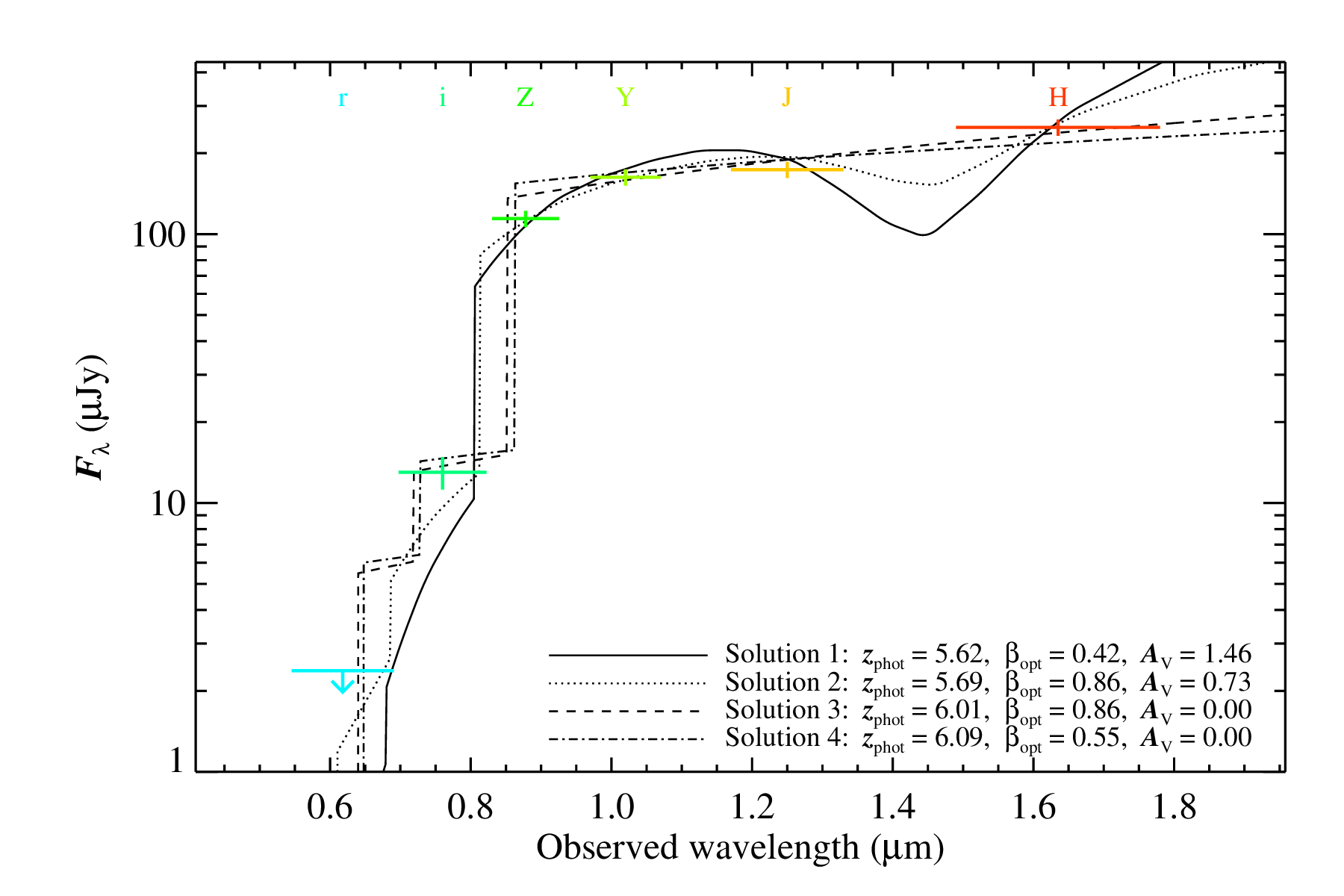}
  \end{center}
  \caption{SED templates  for the four  possible solutions using  a MW
    extinction    law    (see    top    right    panel    of    Figure
    \ref{fig:conts_small}).   The  colored  points correspond  to  the
    measured RATIR photometry, with the filter being marked above each
    measurement.   The black  lines  indicate the  best fits  obtained
    using each  of the  four solutions detailed  in the key  and Table
    \ref{tab:all_fits}.}
  \label{fig:MWseds}
\end{figure*}

The  top right panel  of Figure  \ref{fig:conts_small} shows  the same
parameter space  using the MW  extinction law. In this  instance there
are  four local  maxima in  the prior  weighted probability  map.  The
template SED  for each is  shown in Figure \ref{fig:MWseds}.   The two
higher  $z_{\rm phot}$ solutions  are equivalent  to those  found when
$A_{V}=0$.   There are  two  additional solutions,  however, at  lower
redshift. It  is the shape  of the MW  extinction law that  allows for
these  two solutions,  as  the  2175~\AA\ bump  can  contribute to  the
suppression of  low wavelength  emission.  Even with  the contribution
from this feature, the required  $A_{V}$ is high, with the $\beta_{\rm
  opt}  \sim \beta_{\rm  X}$ solution  requiring $A_{V}=0.73$  and the
lower $\beta_{\rm opt}$ solution requiring $A_{V}=1.46$.\par

An additional point to note is  the selection of the best fit from the
four  possible  MW  solutions.   Table \ref{tab:all_fits}  shows  that
$\chi^{2}$  is   lowest  for  solution  2  ($z_{\rm   phot}  =  5.69$,
$A_{V}=0.73$, $\beta_{\rm  opt} =  0.86$), however, when  weighting by
the  $\beta_{\rm opt}$ prior  probability distribution,  the effective
$\chi^{2}$,    $\chi^{2}_{\rm   eff}$    as   defined    in   Equation
\ref{eq:chisqeff},  is  lower  for  solution 1  ($z_{\rm  phot}=5.62$,
$A_{V}=1.46$,  $\beta_{\rm opt}=0.42$).   When comparing  the multiple
solutions  for  each  extinction  law to  the  obtained  spectroscopic
redshift  of  \citet{2013ApJ...774...26C},  we  find a  solution  with
$\beta_{\rm  opt} \sim  \beta_{\rm  X}$ is  most  consistent with  the
reported value of $z_{\rm spec}=5.913$ in all cases.\par

As an independent test to  determine which of the dust extinction laws
best  represents  the  data,  we  used  the  generic  three  component
extinction    law   measured   in    \citet{1986ApJS...60..305M}   and
\citet{1990ApJS...72..163F,1988ApJ...328..734F,1986ApJ...307..286F}.
Applying    the   prior    probability   distribution    detailed   in
\citet{2001ApJ...553..235R}   yielded  a   fit  where   $z_{\rm  phot}
=6.01_{-0.40}^{+0.13}$,   $\beta_{\rm  opt}=0.88$,   $R_{V}=3.21$  and
$A_{V}=0.00$. Low  dust extinction disfavors  solutions 1 and  2 found
using a  MW type extinction law (see  Table \ref{tab:all_fits}), which
both had the most discrepant values of $z_{\rm phot}$ when compared to
$z_{\rm spec}$ as reported in \citet{2013ApJ...774...26C}.\par

\begin{deluxetable}{ccccccc}
\tablewidth{0pt}
\tabletypesize{\scriptsize}
\tablecaption{Details   of  the   multiple  solutions   to  the   SED  of
    GRB~130606A, stating which host extinction model was used, $z_{\rm
      phot}$,  $\beta_{\rm  opt}$,  $A_{V}$  and $\chi^{2}$  for  each
    model.   These   solutions  were   obtained   using  the   initial
    1-dimensional grid of $z_{\rm grid}$ values.\label{tab:all_fits}}
\tablehead{\colhead{Extinction} & \colhead{Solution} & \colhead{$z_{\rm phot}$} & \colhead{$\beta_{\%rm opt}$} & \colhead{$A_{V}$} & \colhead{$\chi^{2}/ \nu$}
}
\startdata
      None & 1 & 6.01 & 0.86 & 0.00 & 2.55/3 \\
      None & 2 & 6.09 & 0.55 & 0.00 & 5.24/3 \\
      \tableline
      MW & 1 & 5.62 & 0.42 & 1.46 & 2.58/2 \\
      MW & 2 & 5.69 & 0.86 & 0.73 & 2.24/2 \\
      MW & 3 & 6.01 & 0.86 & 0.00 & 2.56/2 \\
      MW & 4 & 6.09 & 0.55 & 0.00 & 5.30/2 \\
      \tableline
      \vspace{0.1cm}
      LMC & 1 & 6.01 & 0.86 & 0.00 & 2.55/2 \\
      LMC & 2 & 5.82 & 0.44 & 0.30 & 4.00/2 \\
      \tableline
      \vspace{0.1cm}
      SMC & 1 & 6.00 & 0.86 & 0.00 & 2.58/2 \\
      SMC & 2 & 5.93 & 0.45 & 0.12 & 3.43/2 \\
\enddata
\vspace{-0.25cm}
\end{deluxetable}

\section{Further tests of the algorithm with simulated RATIR data}
\label{sec:tests}

We constructed  simulated data to  be processed with our  algorithm in
order  to test  the regions  of the  parameter space  in  which robust
values  of $z_{\rm phot}$  could be  recovered from  RATIR photometry.
Specifically, we  wanted to establish  the range of GRB  redshifts for
which  we  could  estimate  $z_{\rm phot}$,  demonstrate  whether  the
algorithm  could  correctly recognize  the  input  extinction law  and
consider the effects of a high amount of dust extinction.\par

\subsection{The range of $z_{\rm phot}$}
\label{sec:test1}

\begin{figure}
  \begin{center}
    \includegraphics[clip,width=8cm]{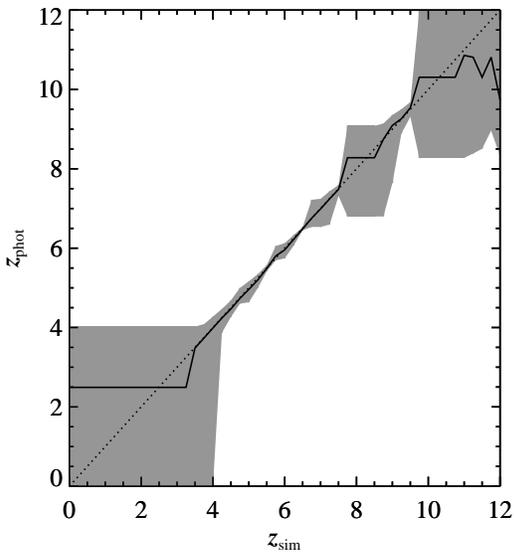}
  \end{center}
  \caption{Fitted $z_{\rm phot}$ as  a function of the input redshift,
    $z_{\rm  sim}$, used  to  create simulated  RATIR magnitudes.  The
    dotted line  denotes equality and the gray  region illustrates the
    3$\sigma$  confidence region  obtained  at each  value of  $z_{\rm
      sim}$. Note $A_{V,{\rm sim}}=0$ in each simulation.}
  \label{fig:ztest1}
\end{figure}

The first tests conducted were simple, with simulated input parameters
$A_{V,{\rm sim}}=0$, $\beta_{\rm opt,sim} = 1.02$ and $0 < z_{\rm sim}
< 12$.   $\beta_{\rm opt,sim}=1.02$ was  chosen as it  corresponded to
the  peak  of  the  \textit{Swift}  distribution  of  $\beta_{\rm  X}$
(obtained   from   the   on-line   repository  at   the   UKSSDC,   see
\citealt{2007A&A...469..379E}). In this  instance we have assumed that
the optical and X-ray parts of  the spectrum lie on the same power-law
slope, such that $\beta_{\rm opt,sim} = \beta_{\rm X,sim}$.  While our
simulated source does  not have dust extinction, we  did allow for the
fitting   routine  to   determine   a  value   for  $A_{V}$.    Figure
\ref{fig:ztest1} shows the fitted values of $z_{\rm phot}$ compared to
$z_{\rm sim}$.\par

Figure  \ref{fig:ztest1} demonstrates  some clear  limitations  of our
algorithm  when applied  to RATIR  data.   First is  the inability  to
measure $z_{\rm phot}\lesssim 4$.  This is physically motivated by the
observed  wavelength  of  Ly$\alpha$.   At  $z_{\rm phot}  =  4$,  the
observed  wavelength of  Ly$\alpha$  is $\sim$  6000  \AA, placing  it
within  the  lower   end  of  the  spectral  coverage   of  the  RATIR
\textit{r}-band.  The  gray region  3$\sigma$ error region  shows that
the  algorithm cannot  discern the  precise value  of  $z_{\rm phot}$,
however  the lack of  an \textit{r}-band  dropout rules  out solutions
where $z_{\rm  phot} >  4$.  Similar analyses  can be  conducted using
GROND  data sets, which  routinely take  simultaneous images  in seven
filters.   Using  the \textit{g'}-band,  GROND  is  able to  determine
$z_{\rm      phot}$       down      to      $z       \approx      2.3$
\citep{2011A&A...526A.153K}.\par

The second limit  is in the $7.5 \lesssim  z_{\rm sim} \lesssim 8.75$.
This is due to the dropout feature lying between the RATIR \textit{Y}-
and \textit{J}-bands.  In these instances, the fitting algorithm tends
towards a single value of  $z_{\rm phot}$. All templates where $z_{\rm
  grid}$  is  such  that   the  SED  features  from  neutral  hydrogen
absorption occur between bands result in an otherwise identical set of
fitted  parameters.   This results  in  fits  which  are of  identical
statistical  merit across  a  range of  $z_{\rm  grid}$. To  calculate
$z_{\rm phot}$ the  fitting routine takes an average  of this range of
$z_{\rm grid}$, rather than simply  taking the value of $z_{\rm grid}$
that normally corresponds to a unique best fit.\par

Despite not having a precise value for $z_{\rm phot}$ when the dropout
occurs  between the  \textit{Y}- and  \textit{J}-bands,  our algorithm
does  still indicate  that it  lies between  the two  bands,  and thus
robustly confirms  it is of  a high redshift.  Such an effect  is also
present when the dropout  occurs between the lower wavelength filters,
however,  the  gap  in  wavelength  coverage is  smaller  and  so  the
corresponding  ranges  of $z_{\rm  phot}$  in  which  this occurs  are
narrower.\par

By  comparison,  GROND  does  not include  a  \textit{Y}-band  filter,
leaving a large gap in  spectral coverage between the \textit{z'}- and
\textit{J}-bands.  Indeed,  \citet{2011A&A...526A.153K} also encounter
a similar effect to that  experienced when using our fitting algorithm
at $7.5  \lesssim z_{\rm sim}  \lesssim 8.75$.  From simulations  of a
comparable nature to those presented here, \citet{2011A&A...526A.153K}
conclude  that the  uncertainty in  their estimate  of  $z_{\rm phot}$
rises from $\Delta  z_{\rm phot} \sim 0.3$ at $z_{\rm  phot} = 6.5$ to
$\Delta z_{\rm  phot} \sim 1.0$ at $z_{\rm  phot}=8.0$.  The 3$\sigma$
confidence region for $z_{\rm  phot}$ as derived from RATIR photometry
similarly rises from $\Delta z_{\rm phot} =0.08$ at $z_{\rm phot}=6.5$
to $\Delta  z_{\rm phot}=2.28$  at $z_{\rm phot}=8.0$.   However, this
increase occurs largely when $7.5 \lesssim z_{\rm phot} \lesssim 8.0$.
Thus, while  both RATIR and  GROND can robustly  infer when a  GRB has
$6.5  \lesssim z_{\rm  phot} \lesssim  8.75$, only  RATIR  can provide
precise  measurements   when  $6.5  \lesssim   z_{\rm  phot}  \lesssim
7.5$.\par

Once $z_{\rm sim}\gtrsim9.5$ we found that the fitting algorithm could
not  recover the  correct value  of  $z_{\rm phot}$.   Aside from  the
dropout  feature once more  falling between  two photometric  bands, a
single  photometric detection in  the \textit{H}-band  is insufficient
for the number of fitted parameters in our model.\par

\subsection{Identifying an extinction law}
\label{sec:test2}

In  cases  where  the  dust  content  is high,  knowing  the  type  of
extinction  law   can  provide  valuable   information  regarding  the
circumburst medium around  the GRB. Thus we tested  the ability of our
algorithm to distinguish between the three \citet{1992ApJ...395..130P}
extinction laws.   To optimize the  effects of the simulated  dust, we
chose  a   redshift,  $z_{\rm  sim}=3.36$.   This   value  places  the
2175~\AA\ bump  at the midpoint  between the RATIR $Z$-  and $Y$-bands
allowing the  SED to  most clearly  define the shape  of the  bump. We
simulated RATIR photometry at $z_{\rm sim}$ with both $A_{V}=0.25$ and
$A_{V}=0.5$,  using the  three  \citet{1992ApJ...395..130P} extinction
laws, in turn.\par

As our algorithm is designed  for use on early-time GRB photometry, to
enable rapid  spectroscopic follow-up, we simulated  the intrinsic GRB
spectra to be  bright, as might be expected during  the first night of
RATIR observations. Each of  the simulated RATIR magnitudes required a
realistic error,  representative of those typically  reported by RATIR
at these epochs, so we assigned  an error of $~\Delta {\rm m}_{\rm AB}=
0.03$ to  each of simulated magnitudes.  In some cases  this error may
prove to  be conservative, with RATIR capable  of measuring magnitudes
to accuracies of $~\Delta {\rm m}_{\rm AB}\sim0.02$ within the first 12
hours      after      the      initial     \textit{Swift}      trigger
\citep{2013GCN..15420...1L}.\par

With  simulated RATIR  photometry in  hand, we  then used  our fitting
algorithm, which  uses templates  including all three  extinction laws
(and an $A_{V}=0$ model), to  find the most representative SED for the
data. Table  \ref{tab:av_comp} shows the full set  of results obtained
when fitting these simulated data.\par

\begin{deluxetable*}{cccccccccc}
\tabletypesize{normalsize}
\tablecolumns{7}
\tablewidth{0pt}
\tablecaption{Recovered  values  of   $A_{V}$,  $z_{\rm  phot}$  and  the
    corresponding $\chi^{2}$ values  for tests conducted with moderate
    ($A_{V,{\rm sim}}=0.25$ and $A_{V}=0.5$) host dust extinction.  In
    all  cases $z_{\rm  sim}=3.36$.  The  type of  extinction  law and
    value of $A_{V}$ used to produce the simulated RATIR photometry are
    shown in the  first and second columns. Th  extinction law used to
    fit $z_{\rm phot}$, the fitted values of $A_{V}$ and $z_{\rm phot}$
    and the associated $\chi^{2}$ value are also shown.\label{tab:av_comp}}
\tablehead{
\colhead{Input Dust} & \colhead{$A_{V{\rm, in}}$} & \colhead{Fitted Dust} & \colhead{$A_{V{\rm ,fit}}$} & \colhead{$z_{\rm phot}$} & \colhead{$\beta_{\rm opt}$} & \colhead{$\chi^{2}/\nu$}
}
    \startdata
    MW & 0.25 & None & 0.00 & 3.65$_{-3.65}^{+0.29}$ & 1.11 & 12.14/3 \\
    MW & 0.25 & MW & 0.24 & 3.36$_{-3.29}^{+0.50}$ & 1.02 & 2.26/2 \\
    MW & 0.25 & LMC & 0.16 & 3.29$_{-3.29}^{+0.56}$ & 1.01 & 7.89/2 \\
    MW & 0.25 & SMC & 1.00 & 0.06$_{-0.06}^{+3.56}$ & 0.53 & 6.57/2 \\
    \tableline
    MW & 0.50 & None & 0.00 & 3.76$_{-3.76}^{+0.25}$ & 1.14 & 40.10/3 \\
    MW & 0.50 & MW & 0.49 & 3.32$_{-0.41}^{+0.32}$ & 1.01 & 0.17/2 \\
    MW & 0.50 & LMC & 0.29 & 3.35$_{-3.35}^{+0.38}$ & 1.02 & 19.28/2 \\
    MW & 0.50 & SMC & 0.78 & 0.16$_{-0.16}^{+3.30}$ & 1.01 & 19.23/2 \\
    \tableline
    LMC & 0.25 & None & 0.00 & 3.99$_{-3.99}^{+0.25}$ & 1.21 & 7.77/3 \\
    LMC & 0.25 & MW & 0.53 & 0.56$_{-0.56}^{+3.22}$ & 1.02 & 0.28/2 \\
    LMC & 0.25 & LMC & 0.30 & 2.19$_{-2.19}^{+1.79}$ & 1.02 & 0.42/2 \\
    LMC & 0.25 & SMC & 0.56 & 0.56$_{-0.56}^{+3.39}$ & 1.03 & 0.48/2 \\
    \tableline
    LMC & 0.50 & None & 0.00 & 4.22$_{-0.25}^{+0.19}$ & 1.37 & 40.12/3 \\
    LMC & 0.50 & MW & 0.96 & 1.11$_{-1.11}^{+0.98}$ & 1.02 & 1.73/2 \\
    LMC & 0.50 & LMC & 0.56 & 3.30$_{-3.30}^{+0.55}$ & 1.02 & 0.58/2 \\
    LMC & 0.50 & SMC & 1.47 & 0.04$_{-0.02}^{+3.77}$ & 1.02 & 1.72/2 \\
    \tableline
    SMC & 0.25 & None & 0.00 & 4.17$_{-4.17}^{+0.14}$ & 1.48 & 14.48/3 \\
    SMC & 0.25 & MW & 0.99 & 0.11$_{-0.11}^{+2.20}$ & 1.02 & 0.44/2 \\
    SMC & 0.25 & LMC & 0.75 & 0.92$_{-0.92}^{+3.00}$ & 1.02 & 0.45/2 \\
    SMC & 0.25 & SMC & 0.73 & 0.91$_{-0.91}^{+3.03}$ & 1.02 & 0.38/2 \\
    \tableline
    SMC & 0.50 & None & 0.00 & 4.41$_{-0.13}^{+0.13}$ & 1.51 & 40.16/3 \\
    SMC & 0.50 & MW & 1.01 & 1.51$_{-1.51}^{+0.53}$ & 1.01 & 0.98/2 \\
    SMC & 0.50 & LMC & 0.96 & 1.73$_{-1.29}^{+2.49}$ & 1.01 & 0.64/2 \\
    SMC & 0.50 & SMC & 0.42 & 3.73$_{-3.29}^{+0.44}$ & 1.02 & 0.15/2 \\
    \enddata
    \vspace{-0.25cm}
\end{deluxetable*}

In all  instances, our algorithm found  the $A_{V}=0$ model  to be the
worst fit to  the simulated photometry. The $\chi^{2}$  value for each
$A_{V}=0$ fit becomes larger with increased $A_{V{\rm, in}}$.\par

For 5  of the 6 test cases,  we found that the  template solution with
the  correct  extinction  law   yielded  the  fit  with  the  smallest
$\chi^{2}$. The margin  by which this fit was  better than alternative
extinction laws  increased notably when  increasing the amount  of dust
from $A_{V}=0.25$ to $A_{V}=0.5$, as expected.\par

\begin{figure*}
  \begin{center}
    \includegraphics[angle=0,clip,width=15cm]{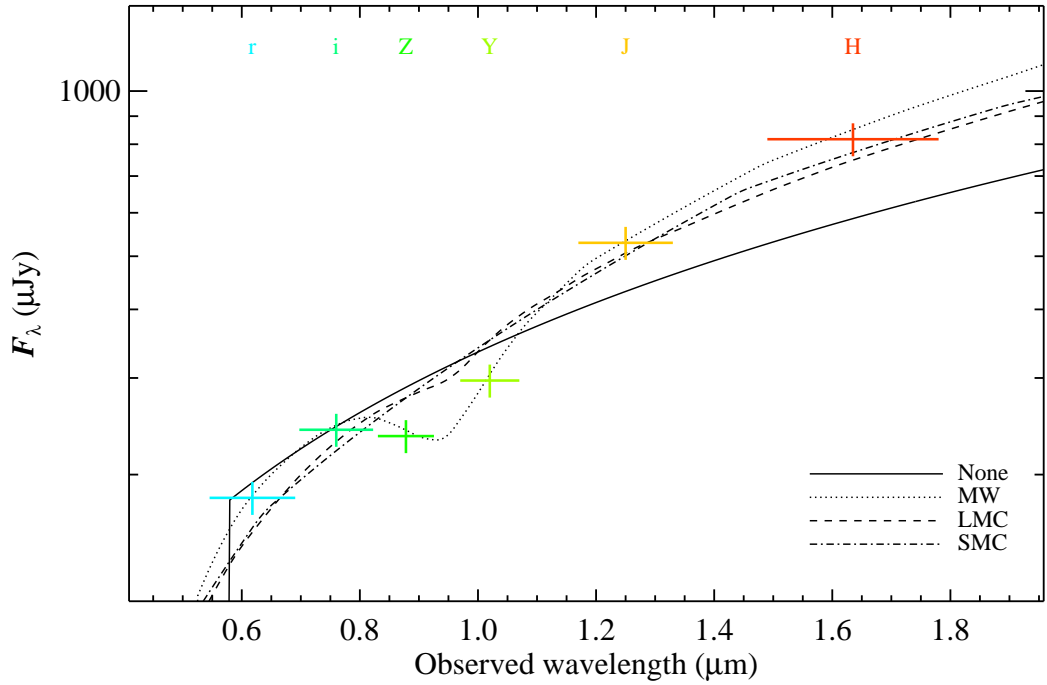}
  \end{center}
  \caption{Fitted  SED templates  to simulated  RATIR  photometry. The
    photometry  was  produced using  a  MW  type  extinction law  with
    $A_{V}=0.5$.  The  black lines each  describe a best  fit template
    obtained  by our  fitting algorithm  using a  different extinction
    law, each being described in the key. The details for each fit are
    available in Table \ref{tab:av_comp}.}
  \label{fig:MWav_type}
\end{figure*}

From the values shown in  Table \ref{tab:av_comp} it is clear that the
MW type extinction law is  the most easily identified by our algorithm,
with the  MW type  template SED having  the lowest $\chi^{2}$  and the
recovered value  of $A_{V}$ most closely matching  $A_{V{\rm, in}}$ in
both  MW tests.  We demonstrate  the SED  templates from  each  of the
extinction   laws    implemented   by   our    algorithm   in   Figure
\ref{fig:MWav_type}.  This more  robust identification  is due  to the
larger prominence of the 2175~\AA\ feature in comparison to the LMC and
SMC extinction laws.\par

The prominence of a strong 2175~\AA\ bump in the MW extinction law has
an  additional  implication for  our  fitting  algorithm.  At  $z_{\rm
  sim}=3.36$, the sharp drop  in flux from neutral hydrogen absorption
is not covered by the RATIR SED. In the absence of large quantities of
dust in the host galaxy along the GRB sight line, our algorithm cannot
determine the  redshift of  the GRB when  $z<4$. However,  occurring at
longer  wavelengths   than  neutral  hydrogen   absorption,  a  strong
2175~\AA\  feature   can  provide  constraints  on   the  galaxy  host
redshift.\par

Table \ref{tab:av_comp}  demonstrates this, as the  recovered value of
$z_{\rm phot}$ is  in good agreement with $z_{\rm sim}$  for a MW type
input extinction law with $A_{V,{\rm in}}=0.5$.  Fitting the resulting
simulated  photometry with  a MW  extinction law  the  3$\sigma$ error
bound states $3.00 < z_{\rm phot} < 3.73$. Thus, in cases of strong MW
type  absorption, our  algorithm  can determine  $z_{\rm  phot}$ to  a
slightly   lower   limit   than   indicated  by   the   results   from
\S~\ref{sec:test1}.    This    is   particularly   of    interest   as
\citet{2012A&A...537A..15S}  note that  the  small sample  of the  GRB
population  with  strong  $A_{V}$  are  best fitted  with  a  MW  type
extinction law.\par

\subsection{Fitting dusty hosts}
\label{sec:test3}

We then considered the case where the intrinsic GRB emission is highly
extinguished  by dust  in  a host  galaxy  at a  variety of  simulated
redshifts.   With  this in  mind,  we  adopted $A_{V,{\rm  sim}}=$0.5.
Taking  into consideration the  capabilities of  the algorithm  in the
$A_{V}=0$ case, we only considered $4 < z_{\rm sim} < 10$. As with the
tests  to identify  dust extinction  we  used a  bright intrinsic  GRB
spectrum, typical of observations within  the first 12 hours after the
\textit{Swift}/BAT  trigger.  Accordingly,  we  assumed  $\Delta  {\rm
  m}_{\rm   AB}  =0.03$   for   all  bands   in   which  we   obtained
detections.\par

\begin{figure*}
  \begin{center}
    \includegraphics[width=7.5cm,clip,angle=0]{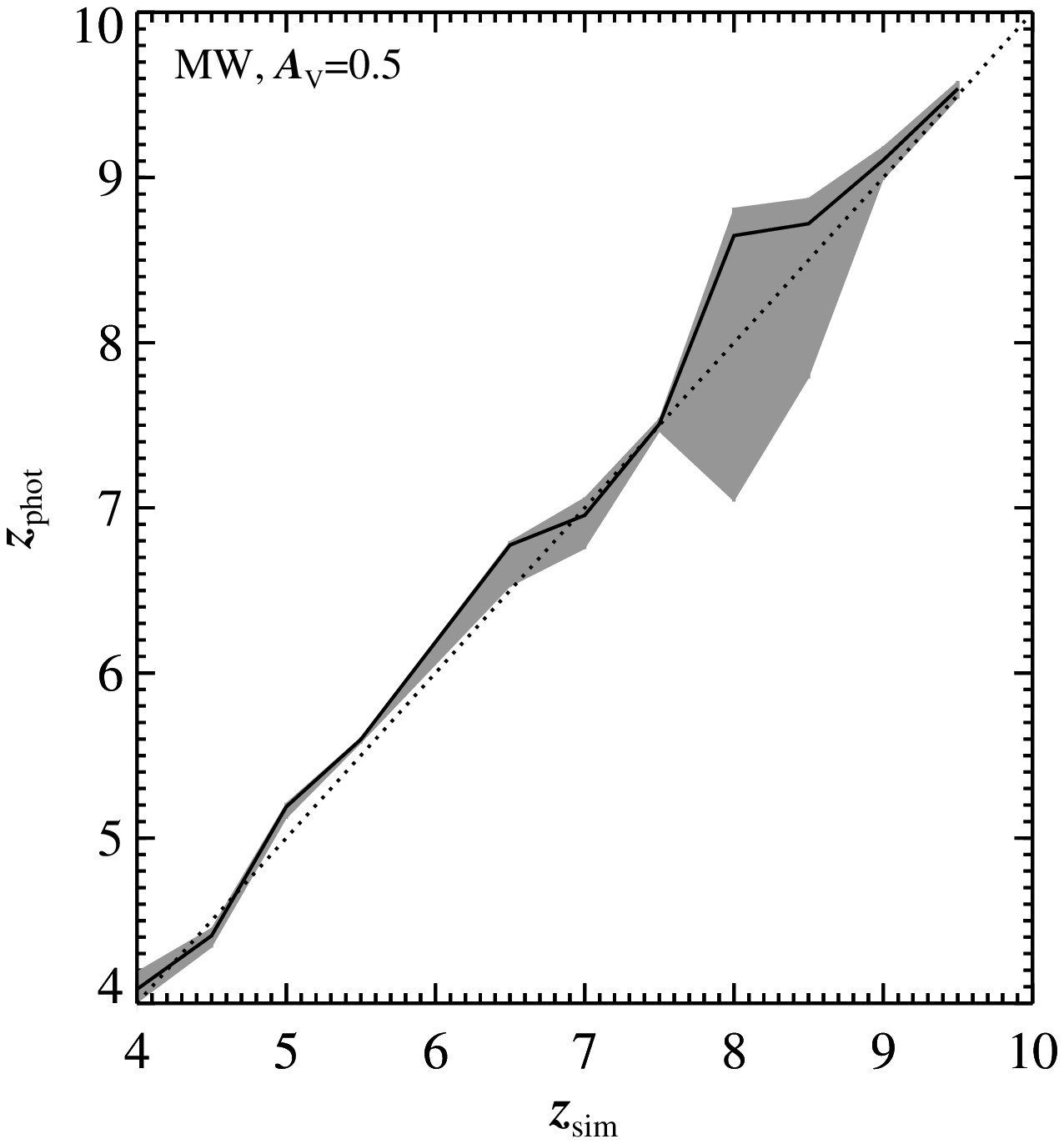} \quad
    \includegraphics[width=7.5cm,clip,angle=0]{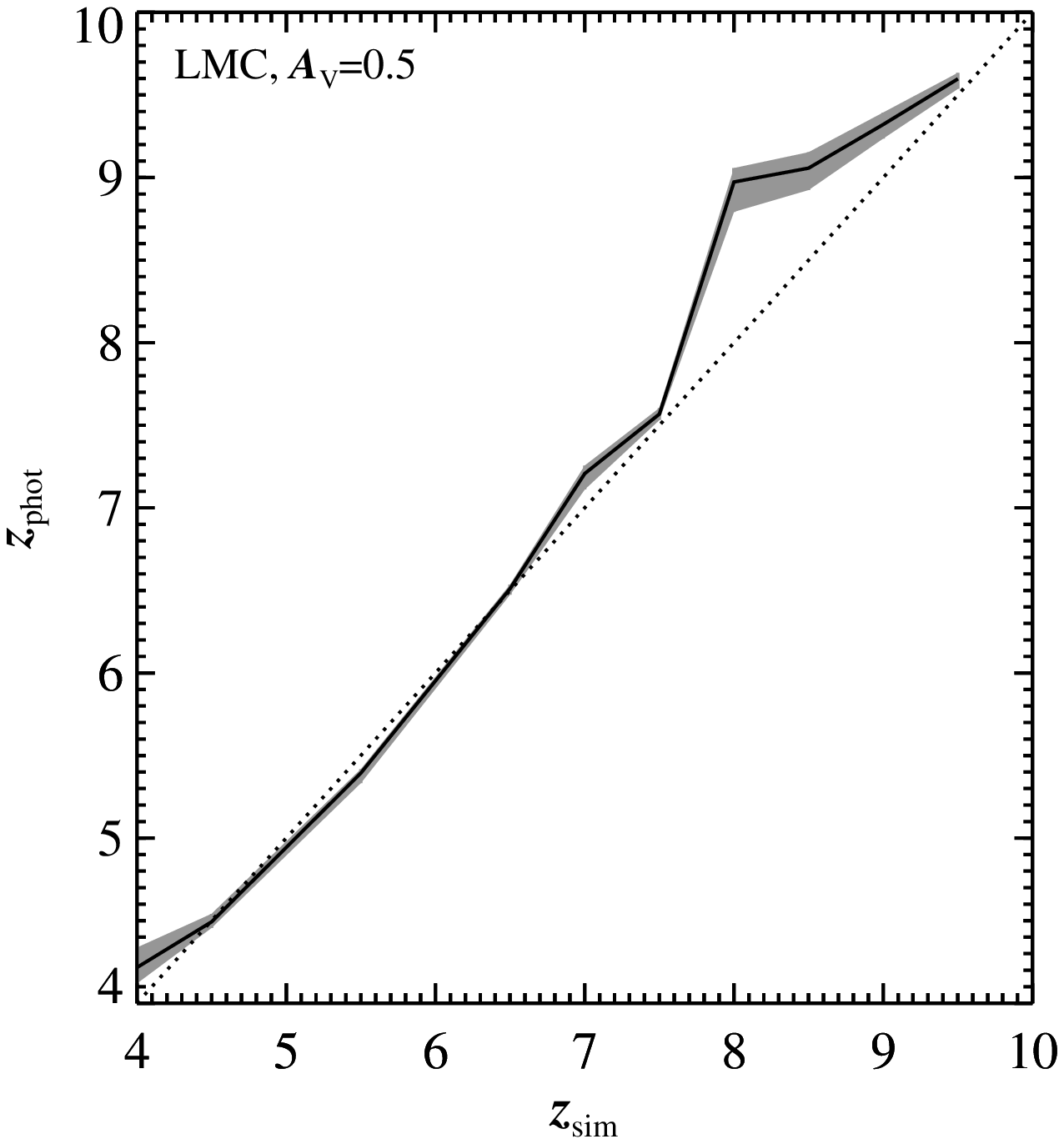}
  \end{center}
  \caption{Fitted $z_{\rm phot}$ as  a function of the input redshift,
    $z_{\rm  sim}$, used  to create  simulated RATIR  magnitudes.  The
    dotted  line denotes  equality.  The left  and  right panels  show
    results from using MW  and LMC type extinction laws, respectively,
    to produce  simulated photometry.   The solid lines  shows $z_{\rm
      phot}$  found for each  $z_{\rm sim}$,  whilst the  gray regions
    illustrates the 3$\sigma$ confidence region about each best fit.}
  \label{fig:ztest2}
\end{figure*}

Figure  \ref{fig:ztest2}  shows the  results  obtained  when the  same
extinction law was used in  both the fitted template and producing the
simulated  photometry. In  each  case  we plot  all  values where  the
template  fitting  algorithm could  successfully  recover  a value  of
$z_{\rm phot}$.  We have only included  plots for the MW  and LMC type
extinction  laws,  as  due  to  the high  $A_{\lambda}$  for  the  SMC
extinction law, the fitting  algorithm was unsuccessful in producing a
unique solution  for most of the  values of $z_{\rm sim}$  with such a
large quantity of SMC type dust in the host galaxy.\par

Figure \ref{fig:ztest2}  show the  recovered values of  $z_{\rm phot}$
remains consistent with $z_{\rm sim}$  prior to $z_{\rm sim} \sim 7.5$
using the MW  and LMC extinction laws. When  $z_{\rm sim} \gtrsim 7.5$
the  algorithm finds  a value  of  $z_{\rm phot}$  that over  predicts
$z_{\rm  sim}$. This  effect is  largely due  to the  neutral hydrogen
absorption  feature falling  between the  RATIR $Y$-  and  $J$- bands,
giving large uncertainty in the precise wavelength at which it occurs.
Even with the larger uncertainty in the MW extinction law simulations,
we can  robustly determine that  $z_{\rm phot}>7.5$ in instances  of a
dusty  GRB sight  line well  represented by  either a  MW or  LMC type
extinction  law. Thus  we retain  the ability  to know  the GRB  is of
interest for further follow-up observations.\par

\section{Conclusions}
\label{sec:conc}

We  have presented  a  template fitting  algorithm  used to  determine
photometric  redshifts from  RATIR data  when a  dropout  candidate is
present. This  algorithm represents the intrinsic GRB  spectrum with a
physically motivated synchrotron  model, includes dust extinction from
the GRB host galaxy and  absorption from intervening clouds of neutral
hydrogen along the  line of sight. Each fitted  SED therefore provides
estimates of $z_{\rm phot}$, $\beta_{\rm opt}$ and $A_{V}$.\par

Applying  this  algorithm  to  the  first  optical  dropout  candidate
observed  by RATIR, GRB~130606A,  we successfully  recover a  value of
$5.6 < z_{\rm phot} < 6.0$  (this is the range of best fit solutions),
with the exact solution being dependent on the extinction law applied.
This  agrees   well  with  the  spectroscopic   redshift  obtained  by
\citet{2013ApJ...774...26C} of $z=5.913$.  Furthermore, we demonstrate
that    using     the    prior    probability     distribution    from
\citet{2001ApJ...553..235R}  for $\beta_{\rm  opt}$  can help  discern
between the multiple potential solutions, as demonstrated particularly
for the MW type extinction law.\par

We also  present the typical  output of the algorithm,  including SEDs
and  probability  maps of  the  parameter  space  for each  extinction
model. Analysis of the finely gridded probability maps, which focus on
the most probably region of $z_{\rm phot}$ $\beta_{\rm opt}$ parameter
space, shows  that for an LMC  or SMC type extinction  law the favored
model  is of  a host  containing  negligible quantities  of dust.  The
degeneracy between  several model solutions with a  MW extinction law
is  partially lifted  by using  the  \citet{2001ApJ...553..235R} prior
probability distribution, which indicates a low value for $A_{V}$ best
fits the  data. With  this prior probability  ruling out  high $A_{V}$
solutions, the best  fit MW solution becomes one  with negligible host
dust content. With this solution,  the three extinction laws all favor
a near identical fit, which is consistent with a zero $A_{V}$ solution
in both the obtained parameter values and $\chi^{2}$ fit statistic.\par

To  ensure our  algorithm is  robust, we  then create  simulated RATIR
photometry, typical  of the quality  expected from the first  night of
observations  after the \textit{Swift}/BAT  trigger.  We  firstly show
that  when there  is no  dust extinction  in the  host galaxy,  we can
successfully recover $z_{\rm phot}$ using the template fitting, in the
ranges $4\lesssim  z_{\rm sim} \lesssim 7.5$  and $8.75\lesssim z_{\rm
  sim} \lesssim 9.5$, for a variety of $\beta_{\rm opt}$ values.  When
$7.5 \lesssim  z_{\rm sim} \lesssim  8.75$ we remain able  to identify
that $z_{\rm phot}$ is within this range, and is therefore a target of
high interest to larger facilities.\par

Introducing a moderate quantity  of dust extinction to our simulations
allows  us to  draw some  conclusions about  the dust  present  in the
host.  MW  type dust  is  the most  easily  identifiable,  due to  the
prominent feature  at 2175~\AA. For weaker amounts  of dust extinction
it  is  difficult to  differentiate  between  LMC  and SMC  extinction
laws. However,  at $A_{V}  \approx 0.5$ the  two can  be distinguished
from one another.\par

An interesting  result from  our tests using  a MW extinction  law and
$A_{V}=0.5$ is that the strong  2175~\AA\ feature in the extinction law
allows  for a  resolved  value  of $z_{\rm  phot}$  even when  neutral
hydrogen  absorption  occurs   slightly  below  the  RATIR  wavelength
coverage. With a  large presence of MW type dust in  a host galaxy our
algorithm can determine  $z_{\rm phot}$ to an improved  lower limit of
$z_{\rm phot}\sim3$.\par

We  also  considered  simulated   data  containing  strong  host  dust
extinction.   When  simulating  RATIR  photometry  with  an  SMC  dust
template,  our algorithm  was unable  to resolve  a solution  for most
values  of  $z_{\rm  sim}$.   This  is  due to  the  large  amount  of
suppression of low wavelength emission from the intrinsic GRB spectrum
by the dust population within the host galaxy. For both the MW and LMC
extinction  laws precise values  of $z_{\rm  phot}$ were  obtained for
$z_{\rm sim} \lesssim  7.5$. After this point the  accuracy of $z_{\rm
  phot}$ reduces, although we remained able to robustly state that the
GRB was of high redshift.  The associated 3$\sigma$ errors were higher
with  the MW  extinction  law due  to  the greater  prominence of  the
2175~\AA\ feature in this type of extinction law. This allows slightly
lower  redshift  solutions  to   be  found  where  the  2175~\AA\ bump
contributes to the suppression of lower wavelength emission.\par

We also note that low mass stars in the Milky Way provide a population
of interlopers that may  occur near a \textit{Swift}/XRT GRB position.
To quantify  the probability  that such a  source is present  in RATIR
observations we considered thick and thin disk components of the Milky
Way      with      exponential      vertical     density      profiles
\citep{2008ApJ...673..864J}.   We  determined  the  local  density  of
M-stars  within a  radius of  20~pc using  the Hipparcos  Main Catalog
\citep{1997A&A...323L..49P}. This volume  was chosen to simultaneously
maximize  completeness and  the  total number  of  sources.  We  cross
referenced  these sources  with  the 2MASS  All-Sky  Catalog of  Point
Sources  \citep{2003tmc..book.....C},  using  the cross-match  service
provided  by   CDS,  Strasbourg,  to  obtain   their  \textit{J}  band
magnitude.   We  found the  chance  probability  that  an M-star  with
\textit{J}$<21$  will be  present  in a  300  square arcsecond  region
centered around  the \textit{Swift}/XRT position is less  than 2.3 \%.
This corresponds  to one (or  fewer) chance M-star in  observations of
approximately  43  different  fields  of  view.  Such  events  can  be
classified as  non-GRB by their  lack of temporal variability  and the
blackbody shape of the SED at low wavelengths.\par

Since December  2012, RATIR has observed  56 GRB fields  of view, with
one instance of an M-star interloper.  RATIR observations of the field
of  GRB~131127A  found  a   red  source  near  the  \textit{Swift}/XRT
position,  suggesting a  high-redshift candidate  ideal  for follow-up
\citep{2013GCN..15525...1B}.   This  source   was  observed  at  later
epochs,    revealing    no    significant    evidence    for    fading
\citep{2013GCN..15555...1B,2013GCN..15548...1I},    leading   to   the
conclusion that this source was not a GRB.\par

The  presented  methodology   is  aimed  specifically  at  identifying
potential high-redshift GRBs, and  providing a preliminary estimate of
redshift. With $z_{\rm phot}$ in hand, the justification of triggering
larger   spectroscopic  facilities  to   measure  a   highly  resolved
spectroscopic  redshift is  significantly higher.  With  the automated
capabilities of RATIR it is therefore possible to request observations
from  such  large  facilities   at  early  epochs  after  the  initial
\textit{Swift}  trigger time,  thereby obtaining  high signal-to-noise
ratio  spectra  required to  answer  fundamental  questions about  the
high-redshift Universe.\par

\acknowledgments

We thank  Sandra Savaglio  for a constructive  referee report  on this
work.  We  thank  the  RATIR   project  team  and  the  staff  of  the
Observatorio   Astron\'{o}mico   Nacional    on   Sierra   San   Pedro
M\'{a}rtir.  RATIR  is  a  collaboration  between  the  University  of
California, the Universidad  Nacional Auton\'{o}ma de M\'{e}xico, NASA
Goddard Space Flight Center,  and Arizona State University, benefiting
from the  loan of an H2RG  detector and hardware  and software support
from Teledyne  Scientific and Imaging.   RATIR, the automation  of the
Harold  L.  Johnson  Telescope  of  the  Observatorio  Astron\'{o}mico
Nacional on Sierra San Pedro M\'{a}rtir, and the operation of both are
funded  through NASA  grants NNX09AH71G,  NNX09AT02G,  NNX10AI27G, and
NNX12AE66G,  CONACyT  grants  INFR-2009-01-122785 and  CB-2008-101958,
UNAM PAPIIT grant IN113810, and UC MEXUS-CONACyT grant CN 09-283.\par

{\it Facilities:} \facility{RATIR}.

\bibliographystyle{apj} 
\bibliography{apj-jour,ms}

\newcommand{\noop}[1]{}
\begin{thebibliography}{}
\expandafter\ifx\csname natexlab\endcsname\relax\def\natexlab#1{#1}\fi

\bibitem[{{Ahn} {et~al.}(2012){Ahn}, {Alexandroff}, {Allende Prieto},
  {Anderson}, {Anderton}, {Andrews}, {Aubourg}, {Bailey}, {Balbinot}, {Barnes},
  \& et~al.}]{2012ApJS..203...21A}
{Ahn}, C.~P., {Alexandroff}, R., {Allende Prieto}, C., {et~al.} 2012, \apjs,
  203, 21

\bibitem[{{Barthelmy} {et~al.}(2005){Barthelmy}, {Barbier}, {Cummings},
  {Fenimore}, {Gehrels}, {Hullinger}, {Krimm}, {Markwardt}, {Palmer},
  {Parsons}, {Sato}, {Suzuki}, {Takahashi}, {Tashiro}, \&
  {Tueller}}]{2005SSRv..120..143B}
{Barthelmy}, S.~D., {Barbier}, L.~M., {Cummings}, J.~R., {et~al.} 2005, \ssr,
  120, 143

\bibitem[{{Becker} {et~al.}(2001){Becker}, {Fan}, {White}, {Strauss},
  {Narayanan}, {Lupton}, {Gunn}, {Annis}, {Bahcall}, {Brinkmann}, {Connolly},
  {Csabai}, {Czarapata}, {Doi}, {Heckman}, {Hennessy}, {Ivezi{\'c}}, {Knapp},
  {Lamb}, {McKay}, {Munn}, {Nash}, {Nichol}, {Pier}, {Richards}, {Schneider},
  {Stoughton}, {Szalay}, {Thakar}, \& {York}}]{2001AJ....122.2850B}
{Becker}, R.~H., {Fan}, X., {White}, R.~L., {et~al.} 2001, \aj, 122, 2850

\bibitem[{{Bertin}(2010)}]{2010ascl.soft10068B}
{Bertin}, E. 2010, {SWarp: Resampling and Co-adding FITS Images Together},
  astrophysics Source Code Library, ascl:1010.068

\bibitem[{{Bertin} \& {Arnouts}(1996)}]{1996A&AS..117..393B}
{Bertin}, E., \& {Arnouts}, S. 1996, \aaps, 117, 393

\bibitem[{{Bolzonella} {et~al.}(2000){Bolzonella}, {Miralles}, \&
  {Pell{\'o}}}]{2000A&A...363..476B}
{Bolzonella}, M., {Miralles}, J.-M., \& {Pell{\'o}}, R. 2000, \aap, 363, 476

\bibitem[{{Burrows} {et~al.}(2005){Burrows}, {Hill}, {Nousek}, {Kennea},
  {Wells}, {Osborne}, {Abbey}, {Beardmore}, {Mukerjee}, {Short}, {Chincarini},
  {Campana}, {Citterio}, {Moretti}, {Pagani}, {Tagliaferri}, {Giommi},
  {Capalbi}, {Tamburelli}, {Angelini}, {Cusumano}, {Br{\"a}uninger}, {Burkert},
  \& {Hartner}}]{2005SSRv..120..165B}
{Burrows}, D.~N., {Hill}, J.~E., {Nousek}, J.~A., {et~al.} 2005, \ssr, 120, 165

\bibitem[{{Butler} {et~al.}(2012){Butler}, {Klein}, {Fox}, {Lotkin}, {Bloom},
  {Prochaska}, {Ramirez-Ruiz}, {de Diego}, {Georgiev}, {Gonz{\'a}lez}, {Lee},
  {Richer}, {Rom{\'a}n}, {Watson}, {Gehrels}, {Kutyrev}, {Bernstein},
  {Alvarez}, {Cese{\~n}a}, {Clark}, {Colorado}, {C{\'o}rdova}, {Farah},
  {Garc{\'{\i}}a}, {Guisa}, {Herrera}, {Lazo}, {L{\'o}pez}, {Luna},
  {Mart{\'{\i}}nez}, {Murillo}, {Murillo}, {N{\'u}{\~n}ez}, {Pedrayes},
  {Quir{\'o}s}, {Ochoa}, {Sierra}, {Moseley}, {Rapchun}, {Robinson}, {Samuel},
  \& {Sparr}}]{2012SPIE.8446E..10B}
{Butler}, N., {Klein}, C., {Fox}, O., {et~al.} 2012, in Society of
  Photo-Optical Instrumentation Engineers (SPIE) Conference Series, Vol. 8446,
  Society of Photo-Optical Instrumentation Engineers (SPIE) Conference Series

\bibitem[{{Butler} {et~al.}(2013{\natexlab{a}}){Butler}, {Watson}, {Kutyrev},
  {Lee}, {Richer}, {Klein}, {Fox}, {Prochaska}, {Bloom}, {Cucchiara}, {Troja},
  {Littlejohns}, {Ramirez-Ruiz}, {de Diego}, {Georgiev}, {Gonzalez},
  {Roman-Zuniga}, {Gehrels}, \& {Moseley}}]{2013GCN..14824...1B}
{Butler}, N., {Watson}, A.~M., {Kutyrev}, A., {et~al.} 2013{\natexlab{a}}, GRB
  Coordinates Network, 14824, 1

\bibitem[{{Butler} {et~al.}(2013{\natexlab{b}}){Butler}, {Watson}, {Kutyrev},
  {Lee}, {Richer}, {Klein}, {Fox}, {Prochaska}, {Bloom}, {Cucchiara}, {Troja},
  {Littlejohns}, {Ramirez-Ruiz}, {de Diego}, {Georgiev}, {Gonzalez},
  {Roman-Zuniga}, {Gehrels}, \& {Moseley}}]{2013GCN..14799...1B}
---. 2013{\natexlab{b}}, GRB Coordinates Network, 14799, 1

\bibitem[{{Butler} {et~al.}(2013{\natexlab{c}}){Butler}, {Watson}, {Kutyrev},
  {Lee}, {Richer}, {Klein}, {Fox}, {Prochaska}, {Bloom}, {Cucchiara}, {Troja},
  {Littlejohns}, {Ramirez-Ruiz}, {de}, {Georgiev}, {Gonzalez}, {Roman-Zuniga},
  {Gehrels}, \& {Moseley}}]{2013GCN..14943...1B}
---. 2013{\natexlab{c}}, GRB Coordinates Network, 14943, 1

\bibitem[{{Butler} {et~al.}(2013{\natexlab{d}}){Butler}, {Watson}, {Kutyrev},
  {Lee}, {Richer}, {Klein}, {Fox}, {Prochaska}, {Bloom}, {Cucchiara}, {Troja},
  {Littlejohns}, {Ramirez-Ruiz}, {de}, {Georgiev}, {Gonzalez}, {Roman-Zuniga},
  {Gehrels}, \& {Moseley}}]{2013GCN..15525...1B}
---. 2013{\natexlab{d}}, GRB Coordinates Network, 15525, 1

\bibitem[{{Butler} {et~al.}(2013{\natexlab{e}}){Butler}, {Watson}, {Kutyrev},
  {Lee}, {Richer}, {Klein}, {Fox}, {Prochaska}, {Bloom}, {Cucchiara}, {Troja},
  {Littlejohns}, {Ramirez-Ruiz}, {de}, {Georgiev}, {Gonzalez}, {Roman-Zuniga},
  {Gehrels}, \& {Moseley}}]{2013GCN..15555...1B}
---. 2013{\natexlab{e}}, GRB Coordinates Network, 15555, 1

\bibitem[{{Butler} {et~al.}(\noop{2014}in prep)}]{2013Butlerprep}
{Butler}, N., {et~al.} \noop{2014}in prep

\bibitem[{{Casali} {et~al.}(2007){Casali}, {Adamson}, {Alves de Oliveira},
  {Almaini}, {Burch}, {Chuter}, {Elliot}, {Folger}, {Foucaud}, {Hambly},
  {Hastie}, {Henry}, {Hirst}, {Irwin}, {Ives}, {Lawrence}, {Laidlaw}, {Lee},
  {Lewis}, {Lunney}, {McLay}, {Montgomery}, {Pickup}, {Read}, {Rees}, {Robson},
  {Sekiguchi}, {Vick}, {Warren}, \& {Woodward}}]{2007A&A...467..777C}
{Casali}, M., {Adamson}, A., {Alves de Oliveira}, C., {et~al.} 2007, \aap, 467,
  777

\bibitem[{{Castro-Tirado} {et~al.}(2013){Castro-Tirado},
  {S{\'a}nchez-Ram{\'{\i}}rez}, {Ellison}, {Jel{\'{\i}}nek},
  {Mart{\'{\i}}n-Carrillo}, {Bromm}, {Gorosabel}, {Bremer}, {Winters},
  {Hanlon}, {Meegan}, {Topinka}, {Pandey}, {Guziy}, {Jeong}, {Sonbas},
  {Pozanenko}, {Cunniffe}, {Fern{\'a}ndez-Mu{\~n}oz}, {Ferrero}, {Gehrels},
  {Hudec}, {Kub{\'a}nek}, {Lara-Gil}, {Mu{\~n}oz-Mart{\'{\i}}nez},
  {P{\'e}rez-Ram{\'{\i}}rez}, {{\v S}trobl}, {{\'A}lvarez-Iglesias},
  {Inasaridze}, {Rumyantsev}, {Volnova}, {Hellmich}, {Mottola}, {Castro
  Cer{\'o}n}, {Cepa}, {G{\"o}{\u g}{\"u}{\c s}}, {G{\"u}ver}, {{\"O}nal Ta{\c
  s}}, {Park}, {Sabau-Graziati}, \& {Tejero}}]{2013arXiv1312.5631C}
{Castro-Tirado}, A.~J., {S{\'a}nchez-Ram{\'{\i}}rez}, R., {Ellison}, S.~L.,
  {et~al.} 2013, ArXiv e-prints, arXiv:1312.5631

\bibitem[{{Chornock} {et~al.}(2013){Chornock}, {Berger}, {Fox}, {Lunnan},
  {Drout}, {Fong}, {Laskar}, \& {Roth}}]{2013ApJ...774...26C}
{Chornock}, R., {Berger}, E., {Fox}, D.~B., {et~al.} 2013, \apj, 774, 26

\bibitem[{{Cucchiara} {et~al.}(2011){Cucchiara}, {Levan}, {Fox}, {Tanvir},
  {Ukwatta}, {Berger}, {Kr{\"u}hler}, {K{\"u}pc{\"u} Yolda{\c s}}, {Wu},
  {Toma}, {Greiner}, {Olivares}, {Rowlinson}, {Amati}, {Sakamoto}, {Roth},
  {Stephens}, {Fritz}, {Fynbo}, {Hjorth}, {Malesani}, {Jakobsson}, {Wiersema},
  {O'Brien}, {Soderberg}, {Foley}, {Fruchter}, {Rhoads}, {Rutledge}, {Schmidt},
  {Dopita}, {Podsiadlowski}, {Willingale}, {Wolf}, {Kulkarni}, \&
  {D'Avanzo}}]{2011ApJ...736....7C}
{Cucchiara}, A., {Levan}, A.~J., {Fox}, D.~B., {et~al.} 2011, \apj, 736, 7

\bibitem[{{Curran} {et~al.}(2008){Curran}, {Wijers}, {Heemskerk}, {Starling},
  {Wiersema}, \& {van der Horst}}]{2008A&A...490.1047C}
{Curran}, P.~A., {Wijers}, R.~A.~M.~J., {Heemskerk}, M.~H.~M., {et~al.} 2008,
  \aap, 490, 1047

\bibitem[{{Cutri} {et~al.}(2003){Cutri}, {Skrutskie}, {van Dyk}, {Beichman},
  {Carpenter}, {Chester}, {Cambresy}, {Evans}, {Fowler}, {Gizis}, {Howard},
  {Huchra}, {Jarrett}, {Kopan}, {Kirkpatrick}, {Light}, {Marsh}, {McCallon},
  {Schneider}, {Stiening}, {Sykes}, {Weinberg}, {Wheaton}, {Wheelock}, \&
  {Zacarias}}]{2003tmc..book.....C}
{Cutri}, R.~M., {Skrutskie}, M.~F., {van Dyk}, S., {et~al.} 2003, {2MASS All
  Sky Catalog of point sources.}

\bibitem[{{Evans} {et~al.}(2007){Evans}, {Beardmore}, {Page}, {Tyler},
  {Osborne}, {Goad}, {O'Brien}, {Vetere}, {Racusin}, {Morris}, {Burrows},
  {Capalbi}, {Perri}, {Gehrels}, \& {Romano}}]{2007A&A...469..379E}
{Evans}, P.~A., {Beardmore}, A.~P., {Page}, K.~L., {et~al.} 2007, \aap, 469,
  379

\bibitem[{{Evans} {et~al.}(2009){Evans}, {Beardmore}, {Page}, {Osborne},
  {O'Brien}, {Willingale}, {Starling}, {Burrows}, {Godet}, {Vetere}, {Racusin},
  {Goad}, {Wiersema}, {Angelini}, {Capalbi}, {Chincarini}, {Gehrels}, {Kennea},
  {Margutti}, {Morris}, {Mountford}, {Pagani}, {Perri}, {Romano}, \&
  {Tanvir}}]{2009MNRAS.397.1177E}
---. 2009, \mnras, 397, 1177

\bibitem[{{Fan} {et~al.}(2006){Fan}, {Strauss}, {Becker}, {White}, {Gunn},
  {Knapp}, {Richards}, {Schneider}, {Brinkmann}, \&
  {Fukugita}}]{2006AJ....132..117F}
{Fan}, X., {Strauss}, M.~A., {Becker}, R.~H., {et~al.} 2006, \aj, 132, 117

\bibitem[{{Fitzpatrick} \& {Massa}(1986)}]{1986ApJ...307..286F}
{Fitzpatrick}, E.~L., \& {Massa}, D. 1986, \apj, 307, 286

\bibitem[{{Fitzpatrick} \& {Massa}(1988)}]{1988ApJ...328..734F}
---. 1988, \apj, 328, 734

\bibitem[{{Fitzpatrick} \& {Massa}(1990)}]{1990ApJS...72..163F}
---. 1990, \apjs, 72, 163

\bibitem[{{Gehrels} {et~al.}(2009){Gehrels}, {Ramirez-Ruiz}, \&
  {Fox}}]{2009ARA&A..47..567G}
{Gehrels}, N., {Ramirez-Ruiz}, E., \& {Fox}, D.~B. 2009, \araa, 47, 567

\bibitem[{{Gehrels} {et~al.}(2004){Gehrels}, {Chincarini}, {Giommi}, {Mason},
  {Nousek}, {Wells}, {White}, {Barthelmy}, {Burrows}, {Cominsky}, {Hurley},
  {Marshall}, {M{\'e}sz{\'a}ros}, {Roming}, {Angelini}, {Barbier}, {Belloni},
  {Campana}, {Caraveo}, {Chester}, {Citterio}, {Cline}, {Cropper}, {Cummings},
  {Dean}, {Feigelson}, {Fenimore}, {Frail}, {Fruchter}, {Garmire}, {Gendreau},
  {Ghisellini}, {Greiner}, {Hill}, {Hunsberger}, {Krimm}, {Kulkarni}, {Kumar},
  {Lebrun}, {Lloyd-Ronning}, {Markwardt}, {Mattson}, {Mushotzky}, {Norris},
  {Osborne}, {Paczynski}, {Palmer}, {Park}, {Parsons}, {Paul}, {Rees},
  {Reynolds}, {Rhoads}, {Sasseen}, {Schaefer}, {Short}, {Smale}, {Smith},
  {Stella}, {Tagliaferri}, {Takahashi}, {Tashiro}, {Townsley}, {Tueller},
  {Turner}, {Vietri}, {Voges}, {Ward}, {Willingale}, {Zerbi}, \&
  {Zhang}}]{2004ApJ...611.1005G}
{Gehrels}, N., {Chincarini}, G., {Giommi}, P., {et~al.} 2004, \apj, 611, 1005

\bibitem[{{Granot} {et~al.}(2000){Granot}, {Piran}, \&
  {Sari}}]{2000ApJ...534L.163G}
{Granot}, J., {Piran}, T., \& {Sari}, R. 2000, \apjl, 534, L163

\bibitem[{{Granot} \& {Sari}(2002)}]{2002ApJ...568..820G}
{Granot}, J., \& {Sari}, R. 2002, \apj, 568, 820

\bibitem[{{Greiner} {et~al.}(2008){Greiner}, {Bornemann}, {Clemens}, {Deuter},
  {Hasinger}, {Honsberg}, {Huber}, {Huber}, {Krauss}, {Kr{\"u}hler},
  {K{\"u}pc{\"u} Yolda{\c s}}, {Mayer-Hasselwander}, {Mican}, {Primak},
  {Schrey}, {Steiner}, {Szokoly}, {Th{\"o}ne}, {Yolda{\c s}}, {Klose}, {Laux},
  \& {Winkler}}]{2008PASP..120..405G}
{Greiner}, J., {Bornemann}, W., {Clemens}, C., {et~al.} 2008, \pasp, 120, 405

\bibitem[{{Greiner} {et~al.}(2009{\natexlab{a}}){Greiner}, {Kr{\"u}hler},
  {Fynbo}, {Rossi}, {Schwarz}, {Klose}, {Savaglio}, {Tanvir}, {McBreen},
  {Totani}, {Zhang}, {Wu}, {Watson}, {Barthelmy}, {Beardmore}, {Ferrero},
  {Gehrels}, {Kann}, {Kawai}, {Yolda{\c s}}, {M{\'e}sz{\'a}ros},
  {Milvang-Jensen}, {Oates}, {Pierini}, {Schady}, {Toma}, {Vreeswijk},
  {Yolda{\c s}}, {Zhang}, {Afonso}, {Aoki}, {Burrows}, {Clemens}, {Filgas},
  {Haiman}, {Hartmann}, {Hasinger}, {Hjorth}, {Jehin}, {Levan}, {Liang},
  {Malesani}, {Pyo}, {Schulze}, {Szokoly}, {Terada}, \&
  {Wiersema}}]{2009ApJ...693.1610G}
{Greiner}, J., {Kr{\"u}hler}, T., {Fynbo}, J.~P.~U., {et~al.}
  2009{\natexlab{a}}, \apj, 693, 1610

\bibitem[{{Greiner} {et~al.}(2009{\natexlab{b}}){Greiner}, {Clemens},
  {Kr{\"u}hler}, {von Kienlin}, {Rau}, {Sari}, {Fox}, {Kawai}, {Afonso},
  {Ajello}, {Berger}, {Cenko}, {Cucchiara}, {Filgas}, {Klose}, {K{\"u}pc{\"u}
  Yolda{\c s}}, {Lichti}, {L{\"o}w}, {McBreen}, {Nagayama}, {Rossi}, {Sato},
  {Szokoly}, {Yolda{\c s}}, \& {Zhang}}]{2009A&A...498...89G}
{Greiner}, J., {Clemens}, C., {Kr{\"u}hler}, T., {et~al.} 2009{\natexlab{b}},
  \aap, 498, 89

\bibitem[{{Greiner} {et~al.}(2011){Greiner}, {Kr{\"u}hler}, {Klose}, {Afonso},
  {Clemens}, {Filgas}, {Hartmann}, {K{\"u}pc{\"u} Yolda{\c s}}, {Nardini},
  {Olivares E.}, {Rau}, {Rossi}, {Schady}, \& {Updike}}]{2011A&A...526A..30G}
{Greiner}, J., {Kr{\"u}hler}, T., {Klose}, S., {et~al.} 2011, \aap, 526, A30

\bibitem[{{Gunn} \& {Peterson}(1965)}]{1965ApJ...142.1633G}
{Gunn}, J.~E., \& {Peterson}, B.~A. 1965, \apj, 142, 1633

\bibitem[{{Hodgkin} {et~al.}(2009){Hodgkin}, {Irwin}, {Hewett}, \&
  {Warren}}]{2009MNRAS.394..675H}
{Hodgkin}, S.~T., {Irwin}, M.~J., {Hewett}, P.~C., \& {Warren}, S.~J. 2009,
  \mnras, 394, 675

\bibitem[{{Hook} {et~al.}(2004){Hook}, {J{\o}rgensen}, {Allington-Smith},
  {Davies}, {Metcalfe}, {Murowinski}, \& {Crampton}}]{2004PASP..116..425H}
{Hook}, I.~M., {J{\o}rgensen}, I., {Allington-Smith}, J.~R., {et~al.} 2004,
  \pasp, 116, 425

\bibitem[{{Im} {et~al.}(2013){Im}, {Choi}, {Lee}, {Ahn}, \&
  {Pak}}]{2013GCN..15548...1I}
{Im}, M., {Choi}, C., {Lee}, H., {Ahn}, H.~N., \& {Pak}, S. 2013, GRB
  Coordinates Network, 15548, 1

\bibitem[{{Juri{\'c}} {et~al.}(2008){Juri{\'c}}, {Ivezi{\'c}}, {Brooks},
  {Lupton}, {Schlegel}, {Finkbeiner}, {Padmanabhan}, {Bond}, {Sesar},
  {Rockosi}, {Knapp}, {Gunn}, {Sumi}, {Schneider}, {Barentine}, {Brewington},
  {Brinkmann}, {Fukugita}, {Harvanek}, {Kleinman}, {Krzesinski}, {Long},
  {Neilsen}, {Nitta}, {Snedden}, \& {York}}]{2008ApJ...673..864J}
{Juri{\'c}}, M., {Ivezi{\'c}}, {\v Z}., {Brooks}, A., {et~al.} 2008, \apj, 673,
  864

\bibitem[{{Kann} {et~al.}(2006){Kann}, {Klose}, \& {Zeh}}]{2006ApJ...641..993K}
{Kann}, D.~A., {Klose}, S., \& {Zeh}, A. 2006, \apj, 641, 993

\bibitem[{{Klein} {et~al.}(2012){Klein}, {Kub{\'a}nek}, {Butler}, {Fox},
  {Kutyrev}, {Rapchun}, {Bloom}, {Farah}, {Gehrels}, {Georgiev},
  {Gonz{\'a}lez}, {Lee}, {Lotkin}, {Moseley}, {Prochaska}, {Ramirez-Ruiz},
  {Richer}, {Robinson}, {Rom{\'a}n-Z{\'u}{\~n}iga}, {Samuel}, {Sparr},
  {Tucker}, \& {Watson}}]{2012SPIE.8453E..2SK}
{Klein}, C.~R., {Kub{\'a}nek}, P., {Butler}, N.~R., {et~al.} 2012, in Society
  of Photo-Optical Instrumentation Engineers (SPIE) Conference Series, Vol.
  8453, Society of Photo-Optical Instrumentation Engineers (SPIE) Conference
  Series

\bibitem[{{Kr{\"u}hler} {et~al.}(2011){Kr{\"u}hler}, {Schady}, {Greiner},
  {Afonso}, {Bottacini}, {Clemens}, {Filgas}, {Klose}, {Koch},
  {K{\"u}pc{\"u}-Yolda{\c s}}, {Oates}, {Olivares E.}, {Page}, {McBreen},
  {Nardini}, {Nicuesa Guelbenzu}, {Rau}, {Roming}, {Rossi}, {Updike}, \&
  {Yolda{\c s}}}]{2011A&A...526A.153K}
{Kr{\"u}hler}, T., {Schady}, P., {Greiner}, J., {et~al.} 2011, \aap, 526, A153

\bibitem[{{Lamb} \& {Reichart}(2000)}]{2000ApJ...536....1L}
{Lamb}, D.~Q., \& {Reichart}, D.~E. 2000, \apj, 536, 1

\bibitem[{{Landsman}(1993)}]{1993ASPC...52..246L}
{Landsman}, W.~B. 1993, in Astronomical Society of the Pacific Conference
  Series, Vol.~52, Astronomical Data Analysis Software and Systems II, ed.
  R.~J. {Hanisch}, R.~J.~V. {Brissenden}, \& J.~{Barnes}, 246

\bibitem[{{Lang} {et~al.}(2010){Lang}, {Hogg}, {Mierle}, {Blanton}, \&
  {Roweis}}]{2010AJ....139.1782L}
{Lang}, D., {Hogg}, D.~W., {Mierle}, K., {Blanton}, M., \& {Roweis}, S. 2010,
  \aj, 139, 1782

\bibitem[{{Liang} \& {Li}(2009)}]{2009ApJ...690L..56L}
{Liang}, S.~L., \& {Li}, A. 2009, \apjl, 690, L56

\bibitem[{{Littlejohns} {et~al.}(2013){Littlejohns}, {Butler}, {Watson},
  {Kutyrev}, {Lee}, {Richer}, {Klein}, {Fox}, {Prochaska}, {Bloom},
  {Cucchiara}, {Troja}, {Ramirez-Ruiz}, {de}, {Georgiev}, {Gonzalez},
  {Roman-Zuniga}, {Gehrels}, \& {Moseley}}]{2013GCN..15420...1L}
{Littlejohns}, O., {Butler}, N., {Watson}, A.~M., {et~al.} 2013, GRB
  Coordinates Network, 15420, 1

\bibitem[{{Madau}(1995)}]{1995ApJ...441...18M}
{Madau}, P. 1995, \apj, 441, 18

\bibitem[{{Massa} \& {Fitzpatrick}(1986)}]{1986ApJS...60..305M}
{Massa}, D., \& {Fitzpatrick}, E.~L. 1986, \apjs, 60, 305

\bibitem[{{M{\'e}sz{\'a}ros}(2006)}]{2006RPPh...69.2259M}
{M{\'e}sz{\'a}ros}, P. 2006, Reports on Progress in Physics, 69, 2259

\bibitem[{{Osborne} {et~al.}(2013){Osborne}, {Beardmore}, {Evans}, \&
  {Goad}}]{2013GCN..14811...1O}
{Osborne}, J.~P., {Beardmore}, A.~P., {Evans}, P.~A., \& {Goad}, M.~R. 2013,
  GRB Coordinates Network, 14811, 1

\bibitem[{{Pei}(1992)}]{1992ApJ...395..130P}
{Pei}, Y.~C. 1992, \apj, 395, 130

\bibitem[{{Perley} {et~al.}(2013){Perley}, {Cenko}, {Corsi}, {Tanvir}, {Levan},
  {Kann}, {Sonbas}, {Wiersema}, {Zheng}, {Zhao}, {Bai}, {Chang}, {Clubb},
  {Frail}, {Fruchter}, {G{\"o}{\u g}{\"u}{\c s}}, {Greiner}, {G{\"u}ver},
  {Horesh}, {Filippenko}, {Klose}, {Mao}, {Morgan}, {Schmidl}, {Stecklum},
  {Tanga}, {Wang}, \& {Xin}}]{2013arXiv1307.4401P}
{Perley}, D.~A., {Cenko}, S.~B., {Corsi}, A., {et~al.} 2013, ArXiv e-prints,
  arXiv:1307.4401

\bibitem[{{Perryman} {et~al.}(1997){Perryman}, {Lindegren}, {Kovalevsky},
  {Hoeg}, {Bastian}, {Bernacca}, {Cr{\'e}z{\'e}}, {Donati}, {Grenon},
  {Grewing}, {van Leeuwen}, {van der Marel}, {Mignard}, {Murray}, {Le Poole},
  {Schrijver}, {Turon}, {Arenou}, {Froeschl{\'e}}, \&
  {Petersen}}]{1997A&A...323L..49P}
{Perryman}, M.~A.~C., {Lindegren}, L., {Kovalevsky}, J., {et~al.} 1997, \aap,
  323, L49

\bibitem[{{Reichart}(2001)}]{2001ApJ...553..235R}
{Reichart}, D.~E. 2001, \apj, 553, 235

\bibitem[{{Roming} {et~al.}(2005){Roming}, {Kennedy}, {Mason}, {Nousek}, {Ahr},
  {Bingham}, {Broos}, {Carter}, {Hancock}, {Huckle}, {Hunsberger}, {Kawakami},
  {Killough}, {Koch}, {McLelland}, {Smith}, {Smith}, {Soto}, {Boyd},
  {Breeveld}, {Holland}, {Ivanushkina}, {Pryzby}, {Still}, \&
  {Stock}}]{2005SSRv..120...95R}
{Roming}, P.~W.~A., {Kennedy}, T.~E., {Mason}, K.~O., {et~al.} 2005, \ssr, 120,
  95

\bibitem[{{Roming} {et~al.}(2009){Roming}, {Koch}, {Oates}, {Porterfield},
  {Vanden Berk}, {Boyd}, {Holland}, {Hoversten}, {Immler}, {Marshall}, {Page},
  {Racusin}, {Schneider}, {Breeveld}, {Brown}, {Chester}, {Cucchiara},
  {DePasquale}, {Gronwall}, {Hunsberger}, {Kuin}, {Landsman}, {Schady}, \&
  {Still}}]{2009ApJ...690..163R}
{Roming}, P.~W.~A., {Koch}, T.~S., {Oates}, S.~R., {et~al.} 2009, \apj, 690,
  163

\bibitem[{{Sari} {et~al.}(1998){Sari}, {Piran}, \&
  {Narayan}}]{1998ApJ...497L..17S}
{Sari}, R., {Piran}, T., \& {Narayan}, R. 1998, \apjl, 497, L17

\bibitem[{{Schady} {et~al.}(2012){Schady}, {Dwelly}, {Page}, {Kr{\"u}hler},
  {Greiner}, {Oates}, {de Pasquale}, {Nardini}, {Roming}, {Rossi}, \&
  {Still}}]{2012A&A...537A..15S}
{Schady}, P., {Dwelly}, T., {Page}, M.~J., {et~al.} 2012, \aap, 537, A15

\bibitem[{{Schmidt} {et~al.}(1989){Schmidt}, {Weymann}, \&
  {Foltz}}]{1989PASP..101..713S}
{Schmidt}, G.~D., {Weymann}, R.~J., \& {Foltz}, C.~B. 1989, \pasp, 101, 713

\bibitem[{{Skrutskie} {et~al.}(2006){Skrutskie}, {Cutri}, {Stiening},
  {Weinberg}, {Schneider}, {Carpenter}, {Beichman}, {Capps}, {Chester},
  {Elias}, {Huchra}, {Liebert}, {Lonsdale}, {Monet}, {Price}, {Seitzer},
  {Jarrett}, {Kirkpatrick}, {Gizis}, {Howard}, {Evans}, {Fowler}, {Fullmer},
  {Hurt}, {Light}, {Kopan}, {Marsh}, {McCallon}, {Tam}, {Van Dyk}, \&
  {Wheelock}}]{2006AJ....131.1163S}
{Skrutskie}, M.~F., {Cutri}, R.~M., {Stiening}, R., {et~al.} 2006, \aj, 131,
  1163

\bibitem[{{Starling} {et~al.}(2007){Starling}, {Wijers}, {Wiersema}, {Rol},
  {Curran}, {Kouveliotou}, {van der Horst}, \&
  {Heemskerk}}]{2007ApJ...661..787S}
{Starling}, R.~L.~C., {Wijers}, R.~A.~M.~J., {Wiersema}, K., {et~al.} 2007,
  \apj, 661, 787

\bibitem[{{Tanvir} {et~al.}(2009){Tanvir}, {Fox}, {Levan}, {Berger},
  {Wiersema}, {Fynbo}, {Cucchiara}, {Kr{\"u}hler}, {Gehrels}, {Bloom},
  {Greiner}, {Evans}, {Rol}, {Olivares}, {Hjorth}, {Jakobsson}, {Farihi},
  {Willingale}, {Starling}, {Cenko}, {Perley}, {Maund}, {Duke}, {Wijers},
  {Adamson}, {Allan}, {Bremer}, {Burrows}, {Castro-Tirado}, {Cavanagh}, {de
  Ugarte Postigo}, {Dopita}, {Fatkhullin}, {Fruchter}, {Foley}, {Gorosabel},
  {Kennea}, {Kerr}, {Klose}, {Krimm}, {Komarova}, {Kulkarni}, {Moskvitin},
  {Mundell}, {Naylor}, {Page}, {Penprase}, {Perri}, {Podsiadlowski}, {Roth},
  {Rutledge}, {Sakamoto}, {Schady}, {Schmidt}, {Soderberg}, {Sollerman},
  {Stephens}, {Stratta}, {Ukwatta}, {Watson}, {Westra}, {Wold}, \&
  {Wolf}}]{2009Natur.461.1254T}
{Tanvir}, N.~R., {Fox}, D.~B., {Levan}, A.~J., {et~al.} 2009, \nat, 461, 1254

\bibitem[{{Tanvir} {et~al.}(2012){Tanvir}, {Levan}, {Fruchter}, {Fynbo},
  {Hjorth}, {Wiersema}, {Bremer}, {Rhoads}, {Jakobsson}, {O'Brien}, {Stanway},
  {Bersier}, {Natarajan}, {Greiner}, {Watson}, {Castro-Tirado}, {Wijers},
  {Starling}, {Misra}, {Graham}, \& {Kouveliotou}}]{2012ApJ...754...46T}
{Tanvir}, N.~R., {Levan}, A.~J., {Fruchter}, A.~S., {et~al.} 2012, \apj, 754,
  46

\bibitem[{{Totani} {et~al.}(2006){Totani}, {Kawai}, {Kosugi}, {Aoki}, {Yamada},
  {Iye}, {Ohta}, \& {Hattori}}]{2006PASJ...58..485T}
{Totani}, T., {Kawai}, N., {Kosugi}, G., {et~al.} 2006, \pasj, 58, 485

\bibitem[{{Totani} {et~al.}(2013){Totani}, {Aoki}, {Hattori}, {Kosugi},
  {Niino}, {Hashimoto}, {Kawai}, {Ohta}, {Sakamoto}, \&
  {Yamada}}]{2013arXiv1312.3934T}
{Totani}, T., {Aoki}, K., {Hattori}, T., {et~al.} 2013, ArXiv e-prints,
  arXiv:1312.3934

\bibitem[{{Ukwatta} {et~al.}(2013){Ukwatta}, {Barthelmy}, {Gehrels}, {Krimm},
  {Malesani}, {Marshall}, {Maselli}, {Melandri}, {Palmer}, \&
  {Stamatikos}}]{2013GCN..14781...1U}
{Ukwatta}, T.~N., {Barthelmy}, S.~D., {Gehrels}, N., {et~al.} 2013, GRB
  Coordinates Network, 14781, 1

\bibitem[{{Updike} {et~al.}(2010){Updike}, {Nicuesa}, {Nardini}, {Kruehler}, \&
  {Greiner}}]{2010GCN..10874...1U}
{Updike}, A., {Nicuesa}, A., {Nardini}, M., {Kruehler}, T., \& {Greiner}, J.
  2010, GRB Coordinates Network, 10874, 1

\bibitem[{{Watson} {et~al.}(2012){Watson}, {Richer}, {Bloom}, {Butler},
  {Cese{\~n}a}, {Clark}, {Colorado}, {C{\'o}rdova}, {Farah}, {Fox-Machado},
  {Fox}, {Garc{\'{\i}}a}, {Georgiev}, {Gonz{\'a}lez}, {Guisa}, {Guti{\'e}rrez},
  {Herrera}, {Klein}, {Kutyrev}, {Lazo}, {Lee}, {L{\'o}pez}, {Luna},
  {Mart{\'{\i}}nez}, {Murillo}, {Murillo}, {N{\'u}{\~n}ez}, {Prochaska},
  {Ochoa}, {Quir{\'o}s}, {Rapchun}, {Rom{\'a}n-Z{\'u}{\~n}iga}, \&
  {Valyavin}}]{2012SPIE.8444E..5LW}
{Watson}, A.~M., {Richer}, M.~G., {Bloom}, J.~S., {et~al.} 2012, in Society of
  Photo-Optical Instrumentation Engineers (SPIE) Conference Series, Vol. 8444,
  Society of Photo-Optical Instrumentation Engineers (SPIE) Conference Series

\bibitem[{{Zafar} {et~al.}(2011){Zafar}, {Watson}, {Tanvir}, {Fynbo},
  {Starling}, \& {Levan}}]{2011ApJ...735....2Z}
{Zafar}, T., {Watson}, D.~J., {Tanvir}, N.~R., {et~al.} 2011, \apj, 735, 2

\end{thebibliography}
\clearpage

\end{document}